\documentclass[twocolumn,showpacs,preprintnumbers,amsmath,amssymb]{revtex4}
\usepackage{graphicx}
\usepackage{dcolumn}
\usepackage{bm}

\newcommand{\as}{a\!\!\!/}
\newcommand{\As}{A\!\!\!/}
\newcommand{\ks}{k\!\!\!/}
\newcommand{\ps}{p\!\!\!/}
\newcommand{\sss}{s\!\!\!/}

\begin{document}

\preprint{}

\title{Mott Scattering of polarized electrons in a circularly polarized laser field }
\author{B. Manaut}\thanks{manaut@fstbm.ac.ma}
\affiliation{ Facult\'e polydisciplinaire, Universit\'e Sultan Moulay Slimane B\'eni Mellal, BP : 523, Maroc.
}
\author{Y. Attaourti}\thanks{attaourti@ucam.ac.ma}
\affiliation{Laboratoire de Physique des Hautes Energies et
d'Astrophysique\\
Facult\'e des Sciences Semlalia, Universit\'e Cadi Ayyad,
Marrakech, Bo\^{\i}te Postale : 2390, Maroc.}
\author{S. Taj}
\affiliation{Laboratoire de Physique des Hautes Energies et
d'Astrophysique\\
Facult\'e des Sciences Semlalia, Universit\'e Cadi Ayyad,
Marrakech, Bo\^{\i}te Postale : 2390, Maroc.}
\author{S. Elhandi}
\affiliation{Laboratoire de Physique des Hautes Energies et
d'Astrophysique\\
Facult\'e des Sciences Semlalia, Universit\'e Cadi Ayyad,
Marrakech, Bo\^{\i}te Postale : 2390, Maroc.}
\begin{abstract}
We present a study of Mott scattering of polarized electrons in the presence of a laser
field with circular polarization using the helicity formalism and the introduction of the well known concept of non
flip differential cross section as well as that of flip differential cross section. The results
we have obtained in the presence of a laser field are coherent with those obtained in the absence
of a laser field. We have studied the relativistic regime as well as the non relativistic regime that are
precisely defined in the text. Two important consistency checks have been carried out successfully. The first one is that the sum
of both differential cross sections (one with spin up, the other with spin down) always gives the unpolarized differential cross section.
The second one is that the relativistic unpolarized differential cross section converges to the non relativistic differential cross
section in the limit of small velocities. Moreover, the results obtained using the sofware Reduce \cite{8} gave rise to non contracted coefficients
that have been dealt with using the geometry chosen.
\end{abstract}
\pacs{34.50.RK, 34.80.Qb, 12.20.Ds}
\maketitle
\section{Introduction}
This work deals simultaneously with three important topics in
Atomic Physics, namely the spin of a Dirac or a Dirac-Volkov
particle, the concept of a spin polarized relativistic particle
and finally how the laser field affects the process of scattering
of such particles. We first give a brief overview of the laser.
The word laser is by now well know to the lay-man because of its
applications in the field of medicine ( surgical and diagnostic
procedures ), telecommunications ( fiber-optic telephone links,
compact disk information storage, etc ) and technology ( laser
drilling of materials, geodesics measurements, newspaper printing
). Lasers come in different shapes, sizes and prices and under
different names as ruby laser ( the first laser to operate \cite
{1} ), helium-neon-argon, semi-conductor laser and others. It is
not possible to explain what a laser is and how it works in a few
words. At this stage, it suffices to say that a laser is a light
source with peculiar characteristics, drastically different from
those of conventional sources such as a candle or a light bulb and
is radiated in a single direction ( not in all directions as in a
light bulb ) which enables any person working in the field to
collect it in a lens and focus it in a region of very small
dimensions. The spectral purity of this process and the
directionality of the laser light dramatically improves the
efficiency of this procedure, making it possible to concentrate a
sizeable amount of power in a small region for different
operations ( like the melting or cutting of metals.) The laser is
basically used as a very powerful light bulb. However they are
others ( like optical communications ), in which its most
important characteristics are the spectral band-width and angular
aperture of the emitted beam. To understand them, one needs to
consider what light is and how it is emitted, which in turn
depends on the emitter, the atom. This task requires the
introduction of some basic concepts of quantum mechanics. In this
work, however, it is possible to describe the laser field
classically since the numbers of photons involved is high \cite
{2}. As for the spin of a particle, according to Landau and
Lifshitz \cite {3}, this property of elementary particles is
peculiar to quantum theory and therefore has in principle no
classical interpretation. In particular, it would be meaningless
to imagine the "intrinsic" angular momentum of an elementary
particle as being the result of its rotation about its own "axis".
It is a fascinating and mysterious complication. Moreover, its
practical effects prevail over the whole of Quantum Physics. The
existence of spin and the statistics associated with it is the
most subtle and ingenious design of Physics, without which the
whole universe will collapse. Spin occupies a unique position
since a wide range of Physics with different degrees of difficulty
is needed for its understanding. The fact is that most textbooks (
even the most advanced ones ) devote at most one chapter to this
subject and give only a utilitarian description. The theory of
special relativity, which is essential for the understanding of
spin and statistics is often forgotten except in deriving the
Dirac equation \cite {4}. The relation between spin and statistics
is hard to understand. Feynman \cite {5} wrote : " it appears to
be one of the few places in Physics which can be stated very
simply, but for which no one has found a simple and easy
explanation. The explanation is down deep in Relativistic Quantum
Mechanics \cite {4}. This probably means that we do not have a
complete understanding of the fundamental principles involved ".
Now, let us turn to the physics of spin polarized free electrons.
Our aim is to oppose the current trend in science that sees
specialists writing primarily for like-minded specialists. It is
for example necessary to give a simple introduction of Mott
Scattering one of the most important techniques in polarized
electron studies. Only in recent years, has it been found possible
to produce electron-beams in which the spins of the particles have
a preferential direction \cite {6}. There are many reasons for the
interest in polarized electrons. One essential reason is that in
physical investigations, one endeavors to define as exactly as
possible the initial and/or the final states of the systems being
considered and also to have electrons in the form of a well
defined beam, that is a beam in which the directions of the
momentum of the individual electrons are as uniform as possible.
Polarization effects in electron scattering were assumed to be
significant only if the electron velocity was comparable to the
velocity of the light. It was not until the $1960$s that large
polarization effects in low-energy electron scattering were
ascertained \cite {7}. It is important, before presenting our
investigation about laser-assisted Mott scattering of polarized
electrons to sketch the principal steps of our treatment. For
purpose of clarity and simplicity, we begin by the most basic
results of Mott Scattering of polarized electrons in the absence
of a laser field. Then, in the presence of a circularly polarized
laser field, we give a detailed account of the formalism used and
we compare the results with those obtained in the absence of a
laser field. The organization of this work is as follows : in
section 2, we present the scattering of polarized electrons by the
Coulomb field of a heavy static charge in the absence of a laser
field and we present the concept of a polarized differential cross
section ( in brief DCS ). We also present the helicity flip DCS as
well as the helicity non flip DCS. Then, we define the degree of
polarization of the scattered electrons. At this stage, it is of a
paramount importance to remark that in experiments, the degree of
polarization of the scattered particles is measured \cite {7}. In
section 3, we present the laser assisted Mott scattering of
polarized electrons in a laser field with circular polarization.
After section 4, devoted to the discussion of the results
obtained, we end by a brief conclusion in section 5. Throughout
this work, we use atomic units $\hbar=m=e=1$ and work with the
metric tensor $g^{\mu\nu}=g_{\mu\nu}=diag(1,-1,-1,-1)$. The
angular frequency we have chosen is $\omega =0.043$ (a.u)$=1.17$
$eV$. It corresponds to the lasing transition of the Nd laser at a
wavelength of $1064$ nm.
\section{Mott Scattering of polarized electrons in the absence of a laser field.}
We begin by one of the simplest process of QED, namely the process
of Mott Scattering of electrons in the lowest order of
perturbation theory. We follow the usual steps of calculations and
we give the transition matrix element corresponding to this
process.
\begin{equation}
S_{fi}=-\frac{i}{c}\int d^{4}x\overline{\psi}_{p_{f}}(x)\As_{coul}
\psi_{p_{i}}
\end{equation}
where
\begin{equation}
\psi_{p_{i}}(x)=\frac{1}{\sqrt{2E_{i}V}}u(p_{i},s_{i})e^{-ip_{i}x}
\end{equation}
describes the incident electron normalized to the volume $V$ and
\begin{equation}
\psi_{p_{f}}(x)=\frac{1}{\sqrt{2E_{f}V}}u(p_{f},s_{f})e^{-ip_{f}x}
\end{equation}
describes the scattered electron with the same volume
normalization. The Coulomb potential $A^0_{coul}(x)$ is generated
by a static heavy nucleus of charge $-Z$, that is :
\begin{equation}
A^{\mu}_{coul}(x)=(-\frac{Z}{|\mathbf{x}|}, 0)
\end{equation}
Therefore, we have for $\As_{coul}(x)$ :
\begin{eqnarray*}
\As_{coul}(x)=\gamma^{0}A_{0}=-\frac{Z\gamma_{0}}{|\mathbf{x}|}
\end{eqnarray*}
The transition matrix element is then given by :
\begin{equation}
S_{fi}=i\frac{Z}{c}\int dx^{0} \int
d^{3}\mathbf{x}\frac{\overline{u}(p_{f},s_{f})}{\sqrt{2E_{f}V}}\frac{\gamma^{0}}{|\mathbf{x}|}\frac{u(p_{i},s_{i})}{\sqrt{2E_{i}V}}
e^{-i(p_{i}-p_{f}).x}
\end{equation}
The integral over $x_{0}$ gives :
\begin{eqnarray}
\int^{+\infty}_{-\infty}dx^{0}
e^{-i(p_{i_{0}}-p_{f_{0}}).x^{0}}&=&
\int^{+\infty}_{-\infty}d(ct)e^{-i(\frac{E_{i}}{c}-\frac{E_{f}}{c}).ct}\nonumber\\
&=& c.2\pi\delta(E_{i}-E_{f})
\end{eqnarray}
while the integral over $\mathbf{x}$ gives :
\begin{equation}
\int d^{3}x e^{i(\mathbf{p}_{i}-\mathbf{p}_{f}).\mathbf{x}}=
\frac{4\pi}{(\mathbf{p}_{i}-\mathbf{p}_{f})^{2}}=\frac{4\pi}{\mathbf{|q|}^{2}}
\end{equation}
where we have introduced the momentum transfer
$\mathbf{q}=\mathbf{p}_{i}-\mathbf{p}_{f}$. Using the standard
procedures of QED \cite {4}, we get for the unpolarized DCS.
\begin{equation}
\frac{d\overline{\sigma}}{d\Omega_{f}}=\frac{Z^{2}}{c^{4}}\frac{1}{|\mathbf{q}|^{4}}\frac{1}{2}\sum_{s_{i},s_{f}}|\overline{u}(p_{f},s_{f})\gamma^{0}
u(p_{i},s_{i})|^{2}\label{a8}
\end{equation}
evaluated for $E_{i}=E_{f}=E$. As a side-result this implies in
turn $|\mathbf{p}_{i}|=|\mathbf{p}_{f}|=|\mathbf{p}|$. In Eq.(\ref{a8})
we have averaged over the initial spin polarizations $s_{i}$ and
summed over the final spin polarizations $s_{f}$. We obtain the
simple result :
\begin{equation}
\frac{1}{2}\sum_{s_{i},s_{f}}|\overline{u}(p_{f},s_{f})\gamma^{0}u(p_{i},s_{i})|^{2}=
2c^{2}[2\frac{E_{i}E_{f}}{c^{2}}-(p_{i}.p_{f})+c^{2}]
\end{equation}
and with $E_{i}=E_{f}=E$ as well as
$|\mathbf{p}_{i}|=|\mathbf{p}_{f}|=|\mathbf{p}|$, we end up with :
\begin{equation}
\frac{1}{2}\sum_{s_{i},s_{f}}|\overline{u}(\mathbf{p}_{f},s_{f})\gamma^{0}u(\mathbf{p}_{i},s_{i})|^{2}=
2c^{2}[\frac{E^{2}}{c^{2}}+\mathbf{|p|}^{2}cos(\theta)+c^{2}]
\end{equation}
where $\theta$ is the scattering angle and
$(p_i.p_f)=E^{2}/c^{2}-\mathbf{p}_{f}.\mathbf{p}_{i}=E^{2}/c^{2}-\mathbf{|p|}^{2}cos(\theta)
$. Using the relation $\beta^{2}E^{2}=c^{2}\mathbf{|p|}^{2}$
($\beta=\frac{v}{c}$) and
$|\mathbf{q}|=|\mathbf{p}_{i}-\mathbf{p}_{f}|=2|\mathbf{p}|\sin{(\theta
/2)}$, the final expression for the unpolarized DCS is :
\begin{eqnarray}
\frac{d\overline{\sigma}}{d\Omega_{f}}_{Mott}&=&
\frac{1}{4}\frac{Z^{2}}{c^{2}\beta^{2}\mathbf{|p|}^{2}}\frac{
1-\beta^{2}\sin^{2}(\theta/2)
}{\sin^{4}(\theta/2)}\nonumber\\
&=&\frac{d\overline{\sigma}}{d\Omega_{f}}_{Ruth}\left(1-\beta^{2}\sin^{2}(\theta/2)\right)
\end{eqnarray}
where
\begin{equation}
\frac{d\overline{\sigma}}{d\Omega_{f}}_{Ruth}=\frac{Z^{2}}{4c^{2}\beta^{2}\mathbf{|p|}^{2}}\frac{1}{\sin^{4}(\theta/2)}
\end{equation}
is the Rutherford unpolarized DCS in the limit of small velocities
($\beta \rightarrow 0$). The calculations carried out up to now
assumed that the initial electron spin is not observed. So, we
focus now on the scattering of polarized Dirac particles. For that
purpose, we need some formalism to describe the spin polarization
\cite {4}. Free electrons with 4-momentum p and spin s are
described by the free spinors $u(p,s)$. The spin 4-vector
$s^{\mu}$ is defined by :
\begin{equation}
s^{\mu}=\frac{1}{c}(|\mathbf{p}|, \frac{E}{c}\hat{p})
\end{equation}
where $\hat{p}=\mathbf{p}/|\mathbf{p}|$ is the unit vector
defining the direction of the 3-vector $\mathbf{p}$. The 4-vector
$s^{\mu}$ is a Lorentz vector in a frame in which the particle
moves with momentum $\mathbf{p}$. The following properties can be
easily checked :
\begin{equation}
s^{\mu}.s_{\mu}=-1
\end{equation}
This is the normalization relation for the 4-vector $s^{\mu}$. One
also have :
\begin{equation}
(p.s)=p^{\mu}.s_{\mu}=0
\end{equation}
It is the orthogonality relation between $p$ and $s$. Next, we
introduce the spin projection operator :
\begin{equation}
\hat\Sigma (s) = \frac{1}{2}(1+\gamma_{5} \sss)
\end{equation}
with
$\gamma_{5}=i\gamma^{0}\gamma^{1}\gamma^{2}\gamma^{3}=-i\gamma_{0}\gamma_{1}\gamma_{2}\gamma_{3}$.
This operator has a simple property :
\begin{equation}
\hat\Sigma (s)\; u(p,+ s)= u(p,+ s)\quad ,\quad \hat\Sigma (s)
\;u(p,- s)=0
\end{equation}
This formalism can be applied to helicity states where the
direction of the spin points along the direction of the momentum
3-vector $\mathbf{p}$.
\begin{equation}
S^{'}_{\lambda}=\lambda \frac{\mathbf{p}}{|\mathbf{p}|} \qquad
\lambda=\pm 1
\end{equation}
Therefore, the definition of a 4-spin vector follows :
\begin{equation}
S^{\mu}_{\lambda}=\frac{\lambda}{c}(|\mathbf{p}|,\frac{E}{c}\hat{p})
\end{equation}
We calculate now the polarized DCS for Mott Scattering of an
electron with a well defined momentum $p_{i}$ and a well defined
spin $s_{i}$. If also the final spin $s_{f}$ is measured, the
polarized DCS reads :
\begin{equation}
\frac{d\sigma}{d\Omega_{f}}=\frac{Z^{2}}{c^{2}}\frac{|\overline{u}(p_{f},s_{f})\gamma^{0}u(p_{i},s_{i})|^{2}}{|\mathbf{q}|^{4}}\label{a20}
\end{equation}
In Eq.(\ref{a20}), one has to be careful since there are no summation on
spin ( either initial or final ) polarizations.\\
We introduce the two operators :
\begin{eqnarray}
\hat\Sigma_{\lambda_i}(s_i)&=\frac{1}{2}(1+\lambda_i\gamma_{5}\sss_{i})\\
\hat\Sigma_{\lambda_f}(s_f)&=\frac{1}{2}(1+\lambda_f\gamma_{5}\sss_{f})
\end{eqnarray}
and we obtain for the polarized DCS :
\begin{eqnarray}
\frac{d\sigma}{d\Omega_{f}}(\lambda_{i},\lambda_{f})=
\frac{Z^{2}}{c^{4}}\frac{1}{4}Tr\{\gamma^{0}(1+\lambda_{i}\gamma_{5}\sss_{i})(\ps_{i}c+c^{2})\nonumber\\
\gamma^{0}(1+\lambda_{f}\gamma_{5}\sss_{f})(\ps_{f}c+c^{2})\}
\end{eqnarray}
using the relations $(p_{i}.s_{f})=\frac{E}{c^{2}}|\mathbf{p}|(1-\cos(\theta))=(p_{f}.s_{i})
$, $(s_i.s_f)=\left(|\mathbf{p}|^2-E^2\cos(\theta)/c^2\right)/c^2$ as well as $(p_{i}.p_{f})=\frac{E^{2}}{c^{2}}-\mathbf{|p|}^{2}\cos(\theta)$\\
we obtain :
\begin{eqnarray}
\frac{d\sigma}{d\Omega_{f}}(\lambda_{i},\lambda_{f})=
\frac{Z^{2}}{|\mathbf{q}^{4}|c^{4}}2\Big\{E^{2}\cos^{2}(\theta/2)+c^{4}\sin^{2}(\theta/2)\nonumber\\
+\lambda_{i}\lambda_{f}(E^{2}\cos^{2}(\theta/2)-c^{4}\sin^{2}(\theta/2))\Big\}
\end{eqnarray}
At this stage, let us note that the $\lambda_{i}$
and $\lambda_{f}$ have the following properties :
\begin{equation}
\lambda_{i}^{2}=\lambda_{f}^{2}=1
\end{equation}
\begin{eqnarray}
\lambda_{i}.\lambda_{f}=1\Longrightarrow\left\{ \begin{array}{c}
                                           \lambda_{i}=\lambda_{f}=1 \\
                                           \lambda_{i}=\lambda_{f}=-1
                                         \end{array}\right.
\end{eqnarray}
\begin{eqnarray}
\lambda_{i}.\lambda_{f}=-1 \Longrightarrow \left\{
\begin{array}{c}
   \lambda_{i}=-\lambda_{f}=1 \\
  \quad \lambda_{i}=-\lambda_{f}=-1
\end{array}\right.
\end{eqnarray}
So, if during the scattering process, $\lambda_{f}=-\lambda_{i}$
which means that a helicity flip occurred, the flip polarized DCS is
:
\begin{eqnarray}
\left.\frac{d\sigma}{d\Omega_{f}}\right|_{flip}&=&\frac{4Z^{2}}{|\mathbf{q}|^{4}c^{4}}c^{4}\sin^2(\theta/2)\nonumber\\
&=&\left.\frac{d\sigma}{d\Omega_{f}}\right|_{Ruth}\frac{c^{4}}{E^{2}}\sin^{2}(\theta/2)
\end{eqnarray}
The case where there is no helicity flip corresponds to
$\lambda_{i}=\lambda_{f}$, so that $\lambda_{i}\lambda_{f}=1$ and
the helicity non flip polarized DCS is given by :
\begin{equation}
\left.\frac{d\sigma}{d\Omega_{f}}\right|_{non
flip}=\left.\frac{d\sigma}{d\Omega_{f}}\right|_{Ruth}\cos^{2}(\theta/2)
\end{equation}
Of course, the sum of the helicity flip polarized DCS and the
helicity non flip polarized DCS must give the unpolarized DCS.
\begin{equation}
\left.\frac{d\overline{\sigma}}{d\Omega_{f}}\right|_{Mott}=\left.\frac{d\sigma}{d\Omega_{f}}\right|_{non
flip}+\left.\frac{d\sigma}{d\Omega_{f}}\right|_{flip}
\end{equation}
Noting that $E^{2}=\mathbf{|p|}^{2}c^{2}+c^{4}$ or
$c^{4}=E^{2}-\mathbf{|p|}^{2}c^{2}$ we have :
\begin{equation}
\frac{c^{4}}{E^{2}}\sin^{2}(\theta/2)+\cos^{2}(\theta/2)=1-\beta^{2}\sin^{2}(\theta/2)
\end{equation}
and the final results is :
\begin{equation}
\left.\frac{d\overline{\sigma}}{d\Omega_{f}}\right|_{Mott}=\left.\frac{d\sigma}{d\Omega_{f}}\right|_{Ruth}\left[1-\beta^{2}\sin^{2}(\theta/2)\right]
\end{equation}
This result is of a paramount importance. \\
For every process, the
sum of the flip polarized DCS and the non flip polarized DCS
always gives the unpolarized DCS. In fact, this is theoretically
exact but, as we shall see very soon (section 3), when it comes to
numerical simulations and where infinite summations over the
number of laser photons are involved, due to a unavoidable lack of
precision, the theoretical exact result becomes a numerical
approximate result.
\begin{figure}[h]
\begin{center}
\includegraphics[angle=0,width=3. in,height=4. in]{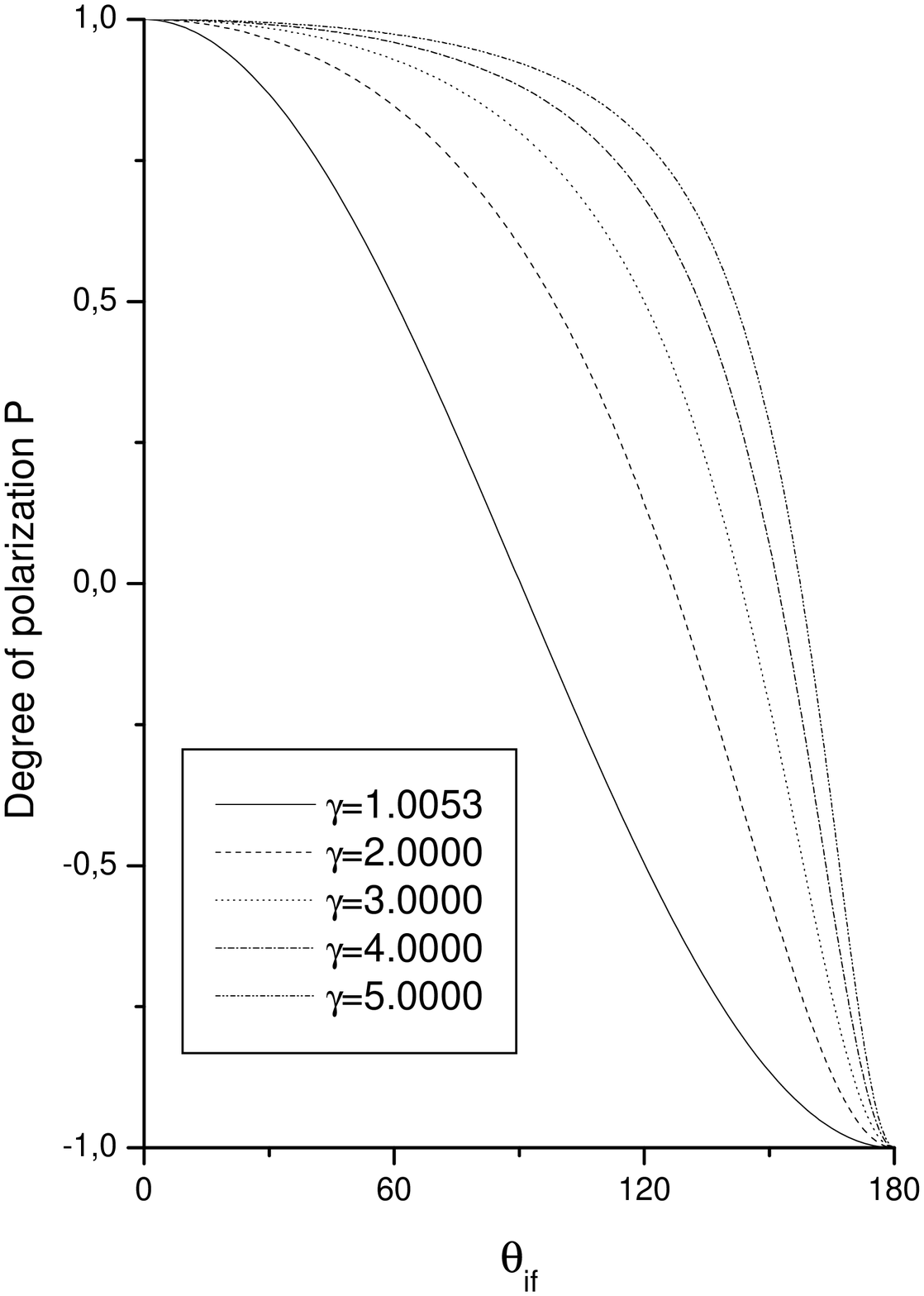}
\caption{Degree of polarization as function of the scattering
angle $\theta_{if} $, for different values of the relativistic
parameter $\gamma $.}
\end{center}
\end{figure}
Another important result is the degree of polarization defined by
:
\begin{equation}
P=\frac{\frac{d\sigma}{d\Omega_{f}}(\lambda_{i}=\lambda_{f}=1)-\frac{d\sigma}{d\Omega_{f}}(\lambda_{i}=-\lambda_{f}=1)}{\frac{d\sigma}{d\Omega_{f}}(\lambda_{i}=\lambda_{f}=1)+\frac{d\sigma}{d\Omega_{f}}(\lambda_{i}=-\lambda_{f}=1)}
\end{equation}
For our process, $P$ reads :
\begin{eqnarray}
P&=&\frac{\cos^{2}(\theta/2)-\frac{c^{4}}{E^{2}}\sin^{2}(\theta/2)}{\cos^{2}(\theta/2)+\frac{c^{4}}{E^{2}}\sin^{2}(\theta/2)}\nonumber\\
&=&1-2\frac{c^{4}\sin^{2}(\theta/2)}{E^{2}\cos^{2}(\theta/2)+c^{4}\sin^{2}(\theta/2)}
\end{eqnarray}
and with $E=\gamma c^{2}$, one has :
\begin{equation}
P=1-2\frac{\sin^{2}(\theta/2)}{\gamma^{2}\cos^{2}(\theta/2)+\sin^{2}(\theta/2)}
\end{equation}
\begin{figure}[h]
\begin{center}
\includegraphics[angle=0,width=3. in,height=4. in]{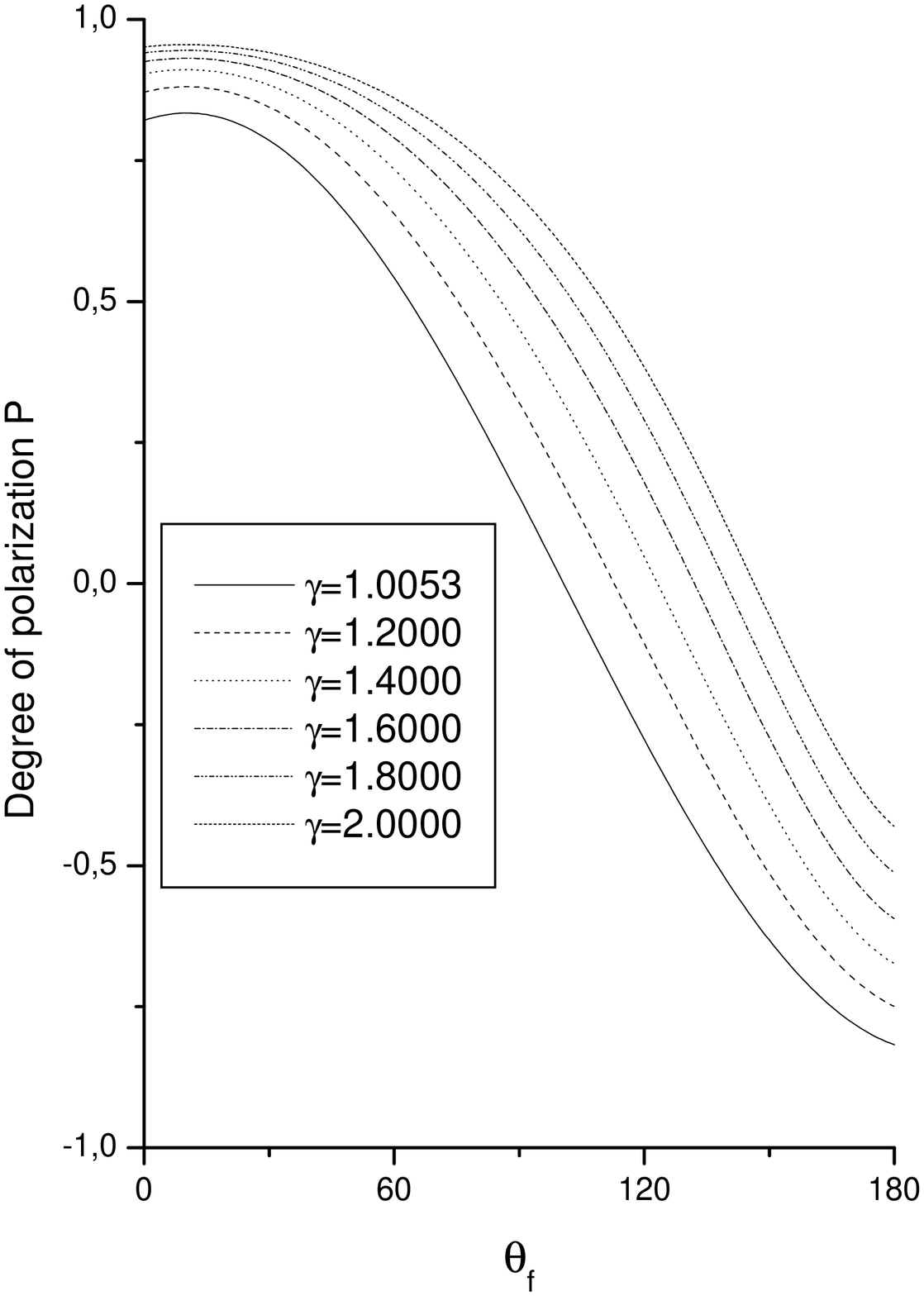}
\caption{ Degree of polarization as function of the angle
$\theta_{f} $, for different values of the relativistic parameter
$\gamma $. }
\end{center}
\end{figure}
We notice that for the process of Mott scattering in the absence
of the laser field, only two parameters are relevant : the
relativistic parameter $\gamma=1/\sqrt{1-\beta^{2}}$ and the
scattering angle $\theta_{if}$. Moreover, one can see from Fig. (1)
that when the collision approaches the relativistic domain, the
degree of polarization becomes less strongly dependent on the
relativistic parameter $\gamma$ and becomes close to a constant
value $P\simeq 1$. The value $P=1$ corresponds to a collision when
no spin flip occurs whereas the value $P=-1$ corresponds to the
reverse situation (there is a spin flip). In Fig. (2), we plot $P$
as a function of the final electron momentum $\theta_{f}$ and a
different picture emerges. Since the scattering angle is not
involved, the minimum value of $P$ shifts to a value greater than
$P=-1$ and this trend is more and more noticeable for increasing
values of $\gamma $.
\section{ Mott scattering of polarized electrons in the presence of a circularly polarized laser field.}
The 4-vector potential $A^{\mu}$ chosen is such that :
\begin{equation}
A^{\mu}=a_{1}^{\mu}\cos(\phi) + a_{2}^{\mu}\sin(\phi)\label{a}
\end{equation}
with $(a_{1}.a_{1})=-\mathbf{|a|}^{2}=-\mathbf{|A|}^{2}$,
$(a_{2}.a_{2})=-\mathbf{|a|}^{2}=-\mathbf{|A|}^{2}$ and the
polarization 4-vectors $a_{1}^{\mu}$ and $a_{2}^{\mu}$ are such
that :
\begin{eqnarray*}
a_{1}^{\mu}=|\mathbf{a}|e_{x}^{\mu}=|\mathbf{a}|(0,1,0,0)
\end{eqnarray*}
and
\begin{eqnarray*}
a_{2}^{\mu}=|\mathbf{a}|e_{y}^{\mu}=|\mathbf{a}|(0,0,1,0)
\end{eqnarray*}
In Eq.(\ref{a}), $\phi=(k.x)=k_{\mu}.x^{\mu}$ with
$k^{\mu}=(k^{0},0,0,k^{3})=\omega/c(1,0,0,1)$. $A^{\mu}$ satisfies
the Lorentz condition :
\begin{equation}
k_{\mu}.A^{\mu}=0
\end{equation}
which implies $(k.a_{1})=(k.a_{2})=0$ i.e, we choose $\mathbf{k}$
along the direction of the $Oz$ axes.\\
Now, in presence of a laser field, the Dirac wave functions
describing the incident and scattered electrons have to be
replaced by the Dirac-Volkov wave function \cite {3} which are
given by :
\begin{equation}
\psi_{q}(x)=R(q)\frac{u(p,s)}{\sqrt{2QV}}e^{iS(q,s)}\label{a40}
\end{equation}
where
\begin{equation}
R(q)=R(p)=1+\frac{\ks.\As}{2c(k.q)}=1+\frac{\ks.\As}{2c(k.p)}
\end{equation}
since the 4-vector $q^{\mu}$ that now replaces $p^{\mu}$ is such
that :
\begin{equation}
q^{\mu}=p^{\mu}-\frac{\overline{A}^{2}}{2c^2(k.p)}k^{\mu}
\end{equation}
The time component of $q^\mu$ is such that :
\begin{eqnarray}
q^{0}=\frac{Q}{c}=\frac{E}{c}-\frac{a^{2}}{2c^2(k.p)}\frac{\omega}{c}=\frac{1}{c}(E-\frac{a^{2}\omega}{2c^2(k.p)})
\end{eqnarray}
We shall see that this 4-vector is the new canonical impulsion
4-vector that the electron feels inside the laser field. The
4-vector $q^{\mu}$ is called the quasi-impulsion and has the
following property :
\begin{eqnarray}
q_{\mu}.q^{\mu}&=&[p-\frac{a^{2}k}{2c^{2}(k.p)}]_{\mu}[p-\frac{a^{2}k}{2c^{2}(k.p)}]^{\mu}\nonumber\\
               &=&p_{\mu}.p^{\mu}-\frac{a^{2}}{c^{2}}\nonumber\\
               &=&c^{2}(1-\frac{a^{2}}{c^{4}})
\end{eqnarray}
Remark also that we can write :
\begin{equation}
R(q)=[1+c(q)\ks\As]=[1+c(p)\ks\As]
\end{equation}
with $c(q)=c(p)=1/[2c(k.q)]=1/[2c(k.p)]$, since $(k.q)=(k.p)$ and $k^2=k^{\mu}.k_{\mu}=0$. The function
$S(q,s)$ appearing in Eq.(\ref{a40}) is such that :
\begin{equation}
S(q,s)=-q.x-i\frac{(p.a_{1})}{c(k.p)}\sin(\phi)+i\frac{(p.a_{2})}{c(k.p)}\cos(\phi)
\end{equation}
One has for the transition matrix element :
\begin{equation}
S_{fi}=-\frac{i}{c}\int
d^{4}x\overline{\psi}_{q_{f}}(x)\As_{coul}(x)\psi_{q_{i}}(x)
\end{equation}
One must be very cautions not to confuse the Coulomb 4-vector
potential due to the charge of the nucleus with the 4-vector
potential of the circularly polarized laser field. As before one
has $\As_{coul}(x)=-Z\gamma^{0}/|\mathbf{x}|$ and therefore :
\begin{equation}
S_{fi}=i\frac{Z}{c}\int
d^{4}x\overline{\psi}_{q_{f}}(x)\frac{\gamma^{0}}{|\mathbf{x}|}\psi_{q_{i}}(x)\label{a48}
\end{equation}
with
\begin{equation}
\psi_{q_{i}}(x)=R(p_{i})\frac{u(p_{i},s_{i})}{\sqrt{2QV}}e^{iS(q_{i},x)}
\end{equation}
and
\begin{equation}
\psi_{q_{f}}(x)=\frac{\overline{u}(p_{f},s_{i})}{\sqrt{2QV}}\overline{R}(p_{f.})e^{-iS(q_{f},x)}
\end{equation}
we transform \quad $e^{iS(q_{i},x)-iS(q_{f},x)}$, if we introduce
:
\begin{equation}
z=\sqrt{\alpha_{1}^{2}+\alpha_{2}^{2}}
\end{equation}
with
\begin{equation}
\alpha_{1}=\frac{1}{c}\left(\frac{(a_{1}.p_{i})}{(k.p_{i})}-\frac{(a_{1}.p_{f})}{(k.p_{f})}\right)
\end{equation}
and
\begin{equation}
\alpha_{2}=\frac{1}{c}\left(\frac{(a_{2}.p_{i})}{(k.p_{i})}-\frac{(a_{2}.p_{f})}{(k.p_{f})}\right)
\end{equation}
we get :
\begin{equation}
S(q_{i},x)-S(q_{f},x)=-(q_{i}-q_{f}).x-z\sin(\phi-\phi_{0})
\end{equation}
where :
\begin{eqnarray}
\cos(\phi_{0})=\frac{\alpha_{1}}{z}\quad \textbf{or} \quad
\sin(\phi_{0})=\frac{\alpha_{2}}{z}
\end{eqnarray}
Now, let us look closely at the quantities in Eq.(\ref{a48}).
First, we have :
\begin{equation}
\overline{\psi}_{q_{f}}(x)\gamma^{0}\psi_{q_{i}}(x)=\overline{u}(p_{f},s_{f})\overline{R}(p_{f})\gamma^{0}R(p_{i})u(p_{i},s_{i})
\end{equation}
In particular, we remark that :
\begin{eqnarray}
\overline{R}(p_{f})\gamma^{0}R(p_{i})=\{1+c(p_{f})[\as_{1}\ks\cos\phi+\as_{2}\ks\sin\phi]\}\nonumber\\
\times\gamma^{0}\{1+c(p_{i})[\ks\as_{1}\cos\phi+\ks\as_{2}\sin\phi]\}\nonumber\\
                                    =\{\gamma^{0}+c(p_{f})[\as_{1}\ks\gamma^{0}\cos\phi+\as_{2}\ks\gamma^{0}\sin\phi]\}\nonumber\\
                                   \times \{1+c(p_{i})[\ks\as_{1}\cos\phi+\ks\as_{2}\sin\phi]\}\label{a57}
\end{eqnarray}
Before invoking the well known
relations involving ordinary Bessel functions, we have first to
transform Eq.(\ref{a57}) in a more compact form. Eq.(\ref{a57})
contains nine terms, and after some algebraic calculations the
final result for
$\overline{R}(q_{f})\gamma^{0}R(q_{i})=\overline{R}(p_{f})\gamma^{0}R(p_{i})$
is :
\begin{eqnarray}
\overline{R}(q_{f})\gamma^{0}R(q_{i})=\gamma^{0}-2\frac{\omega}{c}c(p_{i})c(p_{f})a^{2}\ks\nonumber\\
 +\{c(p_{i})\gamma^{0}\ks
\as_{1}+c(p_{f})\as_{1}\ks\gamma^{0}\}\cos\phi\nonumber\\
 +\{c(p_{i})\gamma^{0}\ks \as_{2}+c(p_{f})\as_{2}\ks\gamma^{0}\}\sin\phi\nonumber\\
=C_0+C_1\cos(\phi)+C_2\sin(\phi)
\end{eqnarray} with
\begin{eqnarray*}
C_0&=&\gamma^{0}-2\frac{\omega}{c}c(p_{i})c(p_{f})a^{2}\ks\\
C_1&=& c(p_{i})\gamma^{0}\ks \as_{1} +c(p_{f})\as_{1}\ks\gamma^{0}\\
C_2&=& c(p_{i})\gamma^{0}\ks \as_{2}+c(p_{f})\as_{2}\ks\gamma^{0}
\end{eqnarray*}
Now, we have :
\begin{eqnarray}
\left\{\begin{array}{c}
1\\
\cos(\phi)\\
\sin(\phi)\end{array}\right\}e^{-iz\sin(\phi-\phi_0)}=\sum_{s=-\infty}^{\infty}\left\{\begin{array}{c}
B_n(z)\\
B_{1n}(z)\\
B_{2n}(z)\end{array}\right\}e^{-in\phi},\label{23}
\end{eqnarray}
with \begin{eqnarray} \left\{\begin{array}{l}
B_n(z)=J_n(z)e^{in\phi_0}\\
B_{1n}(z)=(J_{n+1}(z)e^{i(n+1)\phi_0}+J_{n-1}(z)e^{i(n-1)\phi_0})/2\\
B_{2n}(z)=(J_{n+1}(z)e^{i(n+1)\phi_0}-J_{n-1}(z)e^{i(n-1)\phi_0})/2i
\end{array}\right.
\label{24}
\end{eqnarray}
Therefore :
\begin{eqnarray}
\overline{R}(p_{f})\gamma^{0}R(p_{i})&=&\sum_{n=-\infty}^{+\infty}[C_0B_{n}(z)+C_2B_{1n}(z)\nonumber\\
&&+C_3B_{2n}(z)]e^{-in\phi}\nonumber\\
&=&\sum_{n=-\infty}^{+\infty}\Gamma_n e^{-in\phi}
\end{eqnarray}
The transition matrix element becomes :
\begin{equation}
S_{fi}=iZ2\pi\delta(Q_{i}-Q_{f}+n\omega)\frac{4\pi}{\mathbf{|q|}^{2}}\frac{\overline{u}(p_{f},s_{f})}{\sqrt{2Q_{f}V}}\Gamma_{n}\frac{u(p_{i},s_{i})}{\sqrt{2Q_{i}V}}
\end{equation}
Using the standard procedures of QED, one obtains for the
polarized DCS, we have :
\begin{equation}
\frac{d\sigma}{d\Omega_{f}}=\sum_{n=-\infty}^{+\infty}\frac{d\sigma^{(n)}}{d\Omega_{f}}
\end{equation}
where :
\begin{equation}
\frac{d\sigma^{(n)}}{d\Omega}=\frac{|q_{f}|}{|q_{i}|}\frac{Z^{2}}{c^{4}}\frac{1}{\mathbf{|q|}^{4}}\left|\overline{u}(p_{f},s_{f})\Gamma_{n}u(p_{i},s_{i})\right|^{2}
\end{equation}
evaluated for $Q_{f}=Q_{i}+n\omega$.\\
The squared matrix element is :
\begin{widetext}
\begin{eqnarray}
|\overline{u}(p_{f},s_{f})\Gamma_n
u(p_{i},s_{i})|^{2}&=&u^{\dagger}(p_{i},s_{i})\Gamma^{\dagger}_n\gamma^{0}u(p_{f},s_{f})\overline{u}(p_{f},s_{f})\Gamma_n
u(p_{i},s_{i})\nonumber\\
 &=&\overline{u}(p_{i},s_{i})\gamma^{0}\Gamma^{\dagger}_n\gamma^{0}u(p_{f},s_{f})\overline{u}(p_{f},s_{f})\Gamma_n
                                   u(p_{i},s_{i})
\end{eqnarray}
If the operators $\hat{\Sigma}_{\lambda_i}(s_{i})$ and
$\hat{\Sigma}_{\lambda_f}(s_{f})$ are introduced, we obtain :
\begin{equation}
|\overline{u}(p_{f},s_{f})\Gamma
u(p_{i},s_{i})|^{2}=Tr\left\{\Gamma_n\frac{(1+
\lambda_{i}\gamma_5\sss_{i})}{2}(\ps_{i}c+c^{2})\overline{\Gamma}_n\frac{(1+
\lambda_{f}\gamma_5\sss_{f})}{2}(\ps_{f}c+c^{2})\right\}
\end{equation}
\end{widetext}
where :
\begin{eqnarray}
\overline{\Gamma}_n&=&\gamma^{0}\Gamma^{\dagger}_n\gamma^{0}\nonumber\\
                 &=&[\gamma^{0}-2\frac{\omega}{c}c(p_{f})c(p_{i})a^{2}\ks]B_{n}^{*}\nonumber\\
                 &+&[c(p_{i})\as_{1}\ks\gamma^{0}+c(p_{f})\gamma^{0}\ks \as_{1}]B_{1n}^{*}\nonumber\\
                 &+&[c(p_{i})\as_{2}\ks\gamma^{0}+c(p_{f})\gamma^{0}\ks \as_{2}]B_{2n}^{*}
\end{eqnarray}

\section{RESULTS AND DISCUSSIONS.}

\noindent Before presenting the results and their physical
interpretation, we would like to emphasize the fact that the
REDUCE code we have written gave very long analytical expressions
which were difficult to incorporate in the corresponding FORTRAN
code we wrote to extract figures and tables. From now on, the pair
of indices $n,f$ will stand for " non flip " and the index $f$
alone will stand for " flip ". For the non flip DCS, let us simply
write

\begin{equation}
\left.\frac{d\sigma}{d\Omega}\right|_{n,f}=\left.\sum_{n=-\infty}^{n=\infty}
\frac{d\sigma}{d\Omega}^{(n)}\right|_{n,f}
\end{equation}
where
\begin{eqnarray}
\left.\frac{d\sigma^{(n)}}{d\Omega_{f}}\right|_{(n,f)}=
\frac{|\mathbf{q}_{f}|}{|\mathbf{q_{i}}|}\frac{Z^{2}}{c^{4}}\frac{1}{|\mathbf{q}|^{4}}\Big\{\mathcal{A}(\lambda_{i}=\lambda_{f}=1)J_{n}^{2}(z)\nonumber\\
+ \mathcal{B}(\lambda_{i}=\lambda_{f}=1)[J_{n+1}^{2}(z)+J_{n-1}^{2}(z)]\label{a69}\\
                                                       + \mathcal{C}(\lambda_{i}=\lambda_{f}=1)J_{n+1}(z)J_{n-1}(z)\nonumber\\
                                                       + \mathcal{D}(\lambda_{i}=\lambda_{f}=1)J_{n}(z)[J_{n+1}(z)+J_{n-1}(z)]\Big \}\nonumber
\end{eqnarray}
 The four coefficients $\mathcal{A}(\lambda_{i},\lambda_{f})$, $\mathcal{B}(\lambda_{i},\lambda_{f})$, $\mathcal{C}(\lambda_{i},\lambda_{f})$ and $\mathcal{D}(\lambda_{i},\lambda_{f})$ are very long to
write down so we prefer to focus on their global contents instead
of giving tedious details and explanations. First, we noticed that
in both coefficients
$\mathcal{A}_{n,f}=\mathcal{A}(\lambda_{i}=\lambda_{f}=1)$ and
$\mathcal{C}_{n,f}=\mathcal{C}(\lambda_{i}=\lambda_{f}=1)$, there
is no occurrence of the various completely anti-symmetric tensors
 $\varepsilon_{\alpha\beta\gamma\delta}$ (where the indices are
Lorentz ones and take integer values from 0 to 3). This clearly
means that these tensors were totally contracted and we ended up
with two tractable coefficients that where very easy to
incorporate into the main FORTRAN code. Let us remind the reader
that we used throughout this work, the convention :
\begin{equation}
\varepsilon_{0123}=1
\end{equation}
meaning that $\varepsilon_{\alpha\beta\gamma\delta}=1$ for an even
permutation of the Lorentz indices whereas
$\varepsilon_{\alpha\beta\gamma\delta}=-1$ for an odd permutation of
the Lorentz indices and finally
$\varepsilon_{\alpha\beta\gamma\delta}=0$ otherwise. Second, the
coefficients
$\mathcal{B}_{n,f}=\mathcal{B}(\lambda_{i}=\lambda_{f}=1)$ as well
as $\mathcal{D}_{n,f}=\mathcal{D}(\lambda_{i}=\lambda_{f}=1)$
contained various non contracted tensors. For example in
$\mathcal{B}_{n,f}$, there are thirty one non contracted tensors
involving $\varepsilon_{\alpha\beta\gamma\delta}$ whereas in
$\mathcal{D}_{n,f}$, there are sixty four. Particle physicists are
very often dealing with these. As for us, we were confronted for
the first time with coefficients like for example in
$\mathcal{B}_{n,f}$ :
\begin{equation}
\varepsilon(a_{1},a_{2},k,p_{i})=\varepsilon_{\alpha\beta\gamma\delta}a_{1}^{\alpha}a_{2}^{\beta}k^{\gamma}p_{i}^{\delta}
\end{equation}
and
\begin{equation}
\varepsilon(a_{1},a_{2},k,v)=\varepsilon_{\alpha\beta\gamma\delta}a_{1}^{\alpha}a_{2}^{\beta}k^{\gamma}
v^{\delta}
\end{equation}
\begin{figure}[h]
\begin{center}
\includegraphics[angle=0,width=3. in,height=4. in]{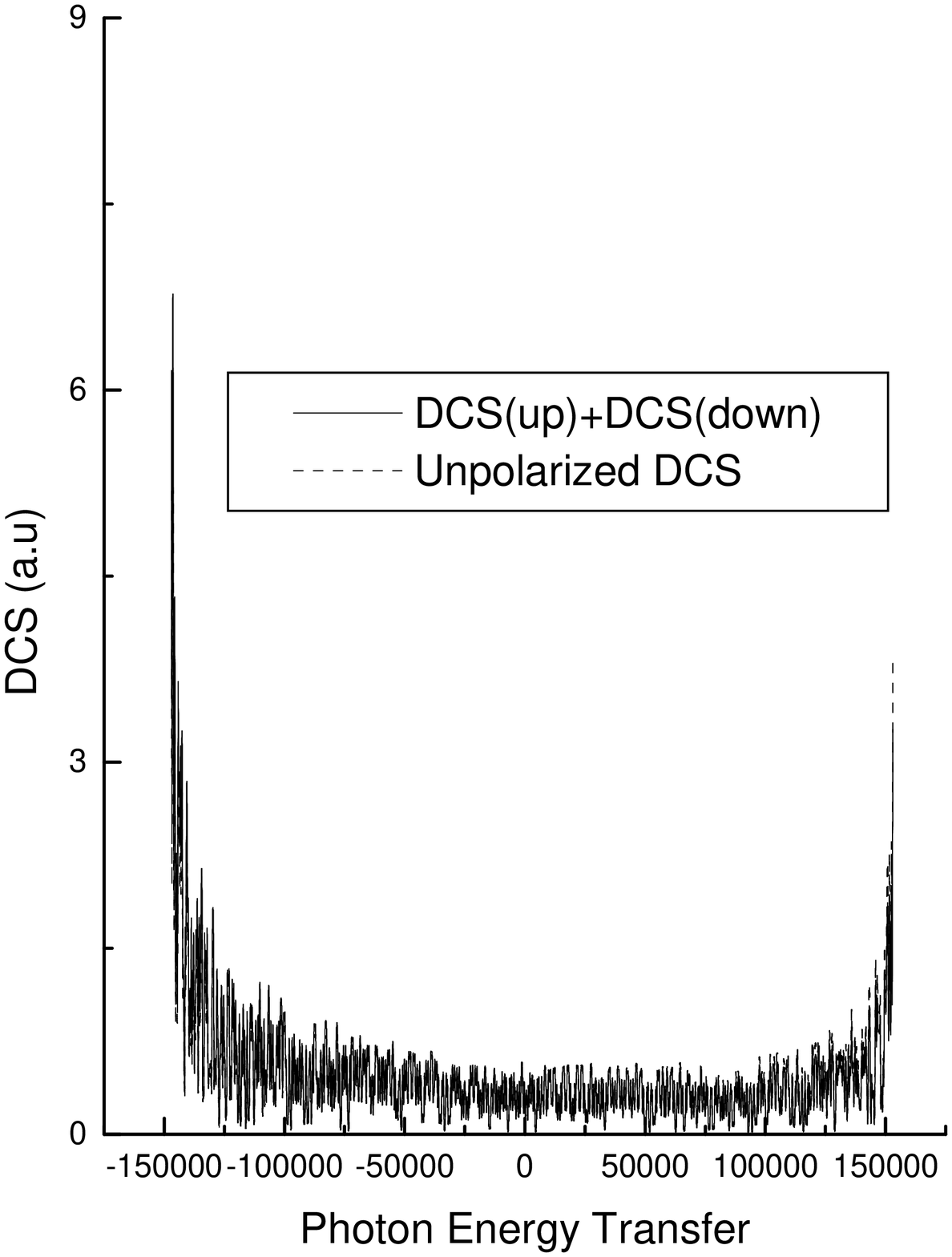}
\caption{Envelope of the relativistic DCS scaled in $10^{-14}$ a.u
as a function of the energy transfer scaled in units of the laser
photon energy $\omega $ for an electrical field strength of
$\mathcal{E}= 1$ a.u and a relativistic parameter $\gamma = 2$.}
\end{center}
\end{figure}
Such coefficients are, at first sight, very complicated to
evaluate. But, when following the conventions of A.G Grozin
\cite{8}, we found that the only non-vanishing contribution of
$a_{1}$ corresponds to the Lorentz index $\alpha=1$ while the only
non-vanishing contribution of $a_{2}$ corresponds to the Lorentz
index $\beta=2$. These helpful values are due to the choice of the
4-potential $A^{\mu}$. Also, it allows to deduce that the only non
trivial indices remaining for $k^{\gamma}$ and $p_{i}^{\delta}$
are the pairs ($\gamma=0$, $\delta=3$) and ($\gamma=3$,
$\delta=0$). So, the non contracted tensor
$\varepsilon(a_{1},a_{2},k,p_{i})$ reduces to :
\begin{eqnarray}
\varepsilon(a_{1},a_{2},k,p_{i})&=&
|\mathbf{a}|^{2}\left[\varepsilon_{1203}k^{0}\mathbf{p}_{i}^{3}+\varepsilon_{1230}k^{3}p_{i}^{0}\right]\nonumber\\
                             &=&|\mathbf{a}|^{2}\frac{\omega}{c}\left[|\mathbf{p}_{i}|\cos\theta_{i}- \frac{E_{i}}{c}\right]
\end{eqnarray}
where $\theta_{i}$ is the angle between the initial electron
momentum and the $Oz$ axis. The coefficient
$\varepsilon(a_{1},a_{2},k,v)$ is easier to evaluate since the
Lorentz index corresponding to $v=(1,0,0,0)$ is zero $\delta=0$. This trick is often used in all QED textbooks \cite{4}. So, the
only choice left for $k$ is $\gamma=3$. Thus, we have :
\begin{eqnarray}
\varepsilon(a_{1},a_{2},k,v)&=&\varepsilon_{1230}a_{1}^{1}a_{2}^{2}k^{3}v^{0}\nonumber\\
                             &=&-|\mathbf{a}|^{2}\frac{\omega}{c}
\end{eqnarray}

It is not necessary to give the whole set of the coefficients
involving the non contracted tensors appearing in
$\mathcal{B}_{n,f}$ and $\mathcal{D}_{n,f}$. It is sufficient to
follow the rules concerning the tensor
$\varepsilon_{\alpha\beta\gamma\delta}$ and the geometry chosen for
$\mathbf{a}_{1}$, $\mathbf{a}_{2}$ and $\mathbf{k}$, bearing in
mind that
$a_{1}^{\mu}=|\mathbf{a}|e_{1}^{\mu}=|\mathbf{a}|(0,1,0,0)$,
$a_{2}^{\nu}=|\mathbf{a}|e_{1}^{\nu}=|\mathbf{a}|(0,0,1,0)$ and
finally $k^{\sigma}=\frac{\omega}{c}(1,0,0,1)$. The same holds for
the coefficients $\mathcal{B}_{f}=\mathcal{B}(\lambda_{i}=-\lambda_{f}=1)$ and
$\mathcal{D}_{f}=\mathcal{D}(\lambda_{i}=-\lambda_{f}=1)$. The complete expressions of
the coefficients $\mathcal{A}$, $\mathcal{B}$, $\mathcal{C}$ and
$\mathcal{D}$ in both cases $(nf,f)$ can explicitly be found in the Appendix.
It is important to remind that these coefficients occur only in
presence of a laser field. If we put $A^{\mu}=(0,\mathbf{0})$, we
easily recover the results of section 2. Now, let us turn to the
discussion of the results concerning
$\left.d\sigma/d\Omega_{f}\right|_{n,f}$ and
$\left.d\sigma/d\Omega_{f}\right|_{f}$. We chose two regimes :
 a) the relativistic regime corresponding to a relativistic parameter
$\gamma=2$ and an electric field strength $\mathcal{E}=1$ a.u and
b) the non relativistic regime corresponding to a relativistic
parameter $\gamma=1.0053$ and an electric field strength
$\mathcal{E}=0.05$ a.u. There are two consistency checks that must
be made. First, we have to show numerically that the sum of
$(DCS)_{f}$ and $(DCS)_{n,f}$ always gives the unpolarized DCS and
second, the non relativistic description for the unpolarized DCS
must give the unpolarized DCS we have found in the formalism we
have developed.
\subsection{The relativistic regime}
\begin{figure}[h]
\begin{center}
\includegraphics[angle=0,width=3. in,height=4. in]{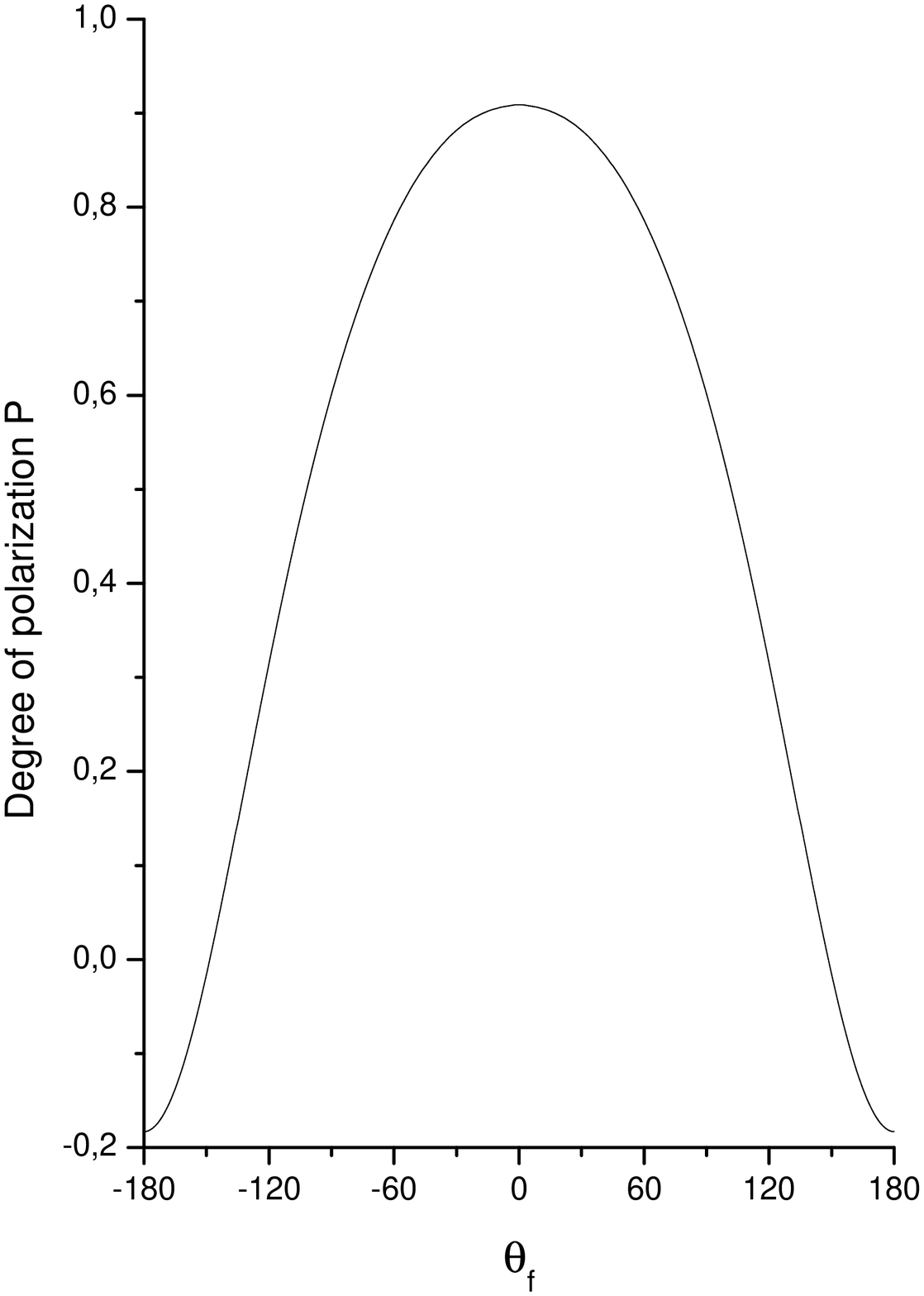}
\caption{Degree of polarization $P$ for an electrical field
strength of $\mathcal{E}= 1.$ a.u and a relativistic parameter
$\gamma = 2.$}
\end{center}
\end{figure}
\noindent This regime corresponds to a total energy of the
incoming electron $E_i=0.510858\;MeV$  which is the rest
energy of the electron so, it is located below the threshold
energy needed for the process of pair creation. As for the
cutoffs, the negative part of the spectrum corresponds
approximately to $-150000$ photons (emitted by the laser). As a
side-result, the various DCS ($n,f$ or $f$) decrease abruptly
because the arguments of the Bessel functions are close to their
indices. Once again, there is an asymmetry between the emission
part of the envelope and the absorption part of the same envelope
for the unpolarized DCSs because of the presence of the
denominator $|\mathbf{q}|^{4}$ in Eq.(\ref{a69}). Now, the positive part
of the spectrum corresponds approximately to $175000$ photons
absorbed by the laser field. In Fig. (3), we show the envelope of
$d\sigma/d\Omega_{f}$ as a function of the net number of photons
exchanged. The cutoffs are $n\simeq -150000$ photons for the
negative part of the envelope and $n_{max}\simeq 170000$ photons
for the positive part of the envelope that corresponds to the absorptive part of the spectrum. The geometry chosen for
Fig. (3) is $\theta_{i}=45^{\circ}$, $\phi_{i}=0^{\circ}$,
$\theta_{f}=50^{\circ}$ and $\phi_{f}=90^{\circ}$. One many ask
legitimately if such cutoffs are geometry-dependent. The answer is
that indeed they are geometry dependent since for a different
choice of the initial, final momentum angular coordinates
$\mathbf{p}_{i}$ and $\mathbf{p}_{f}$ namely
$\theta_{i}=30^{\circ}$, $\phi_{i}=0^{\circ}$,
$\theta_{i}=75^{\circ}$ and $\phi_{f}=90^{\circ}$ the cutoffs
obtained are $n_{min}\simeq -147000$ photons and $n_{max}\simeq
+154000$ photons. In Fig. (4), we show the degree of polarization $P$ corresponding to this regime.

\begin{figure}[h]
\begin{center}
\includegraphics[angle=0,width=3. in,height=4. in]{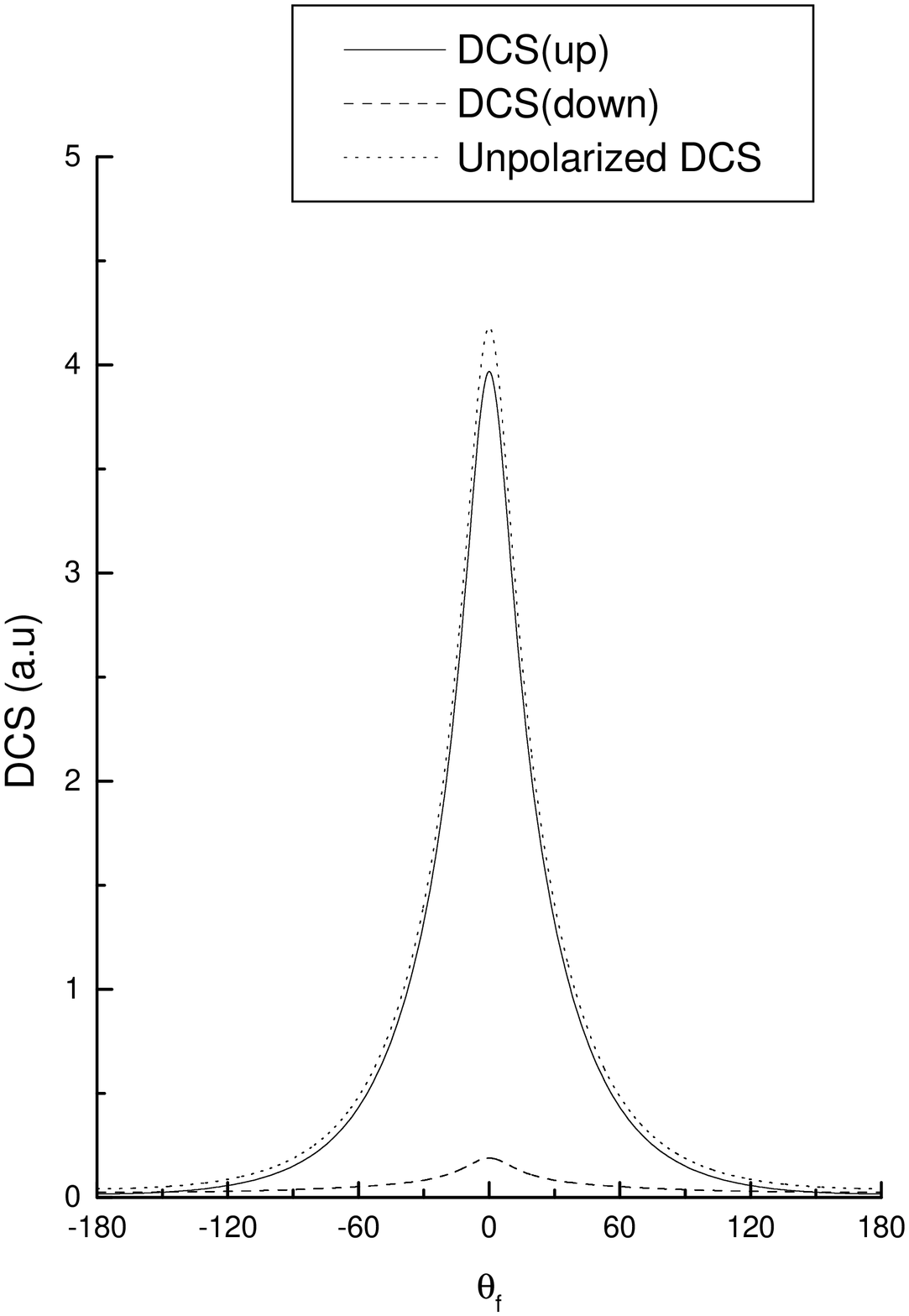}
\caption{The various relativistic DCSs scaled in $10^{-11}$ a.u as
a function of the angle $\theta_{f}$ in degrees for an electrical
field strength of $\mathcal{E}= 1.$ a.u and a relativistic
parameter $\gamma = 2$. The corresponding number of photons
exchanged is $\pm 500$.}
\end{center}
\end{figure}
\begin{figure}[h]
\begin{center}
\includegraphics[angle=0,width=3. in,height=4. in]{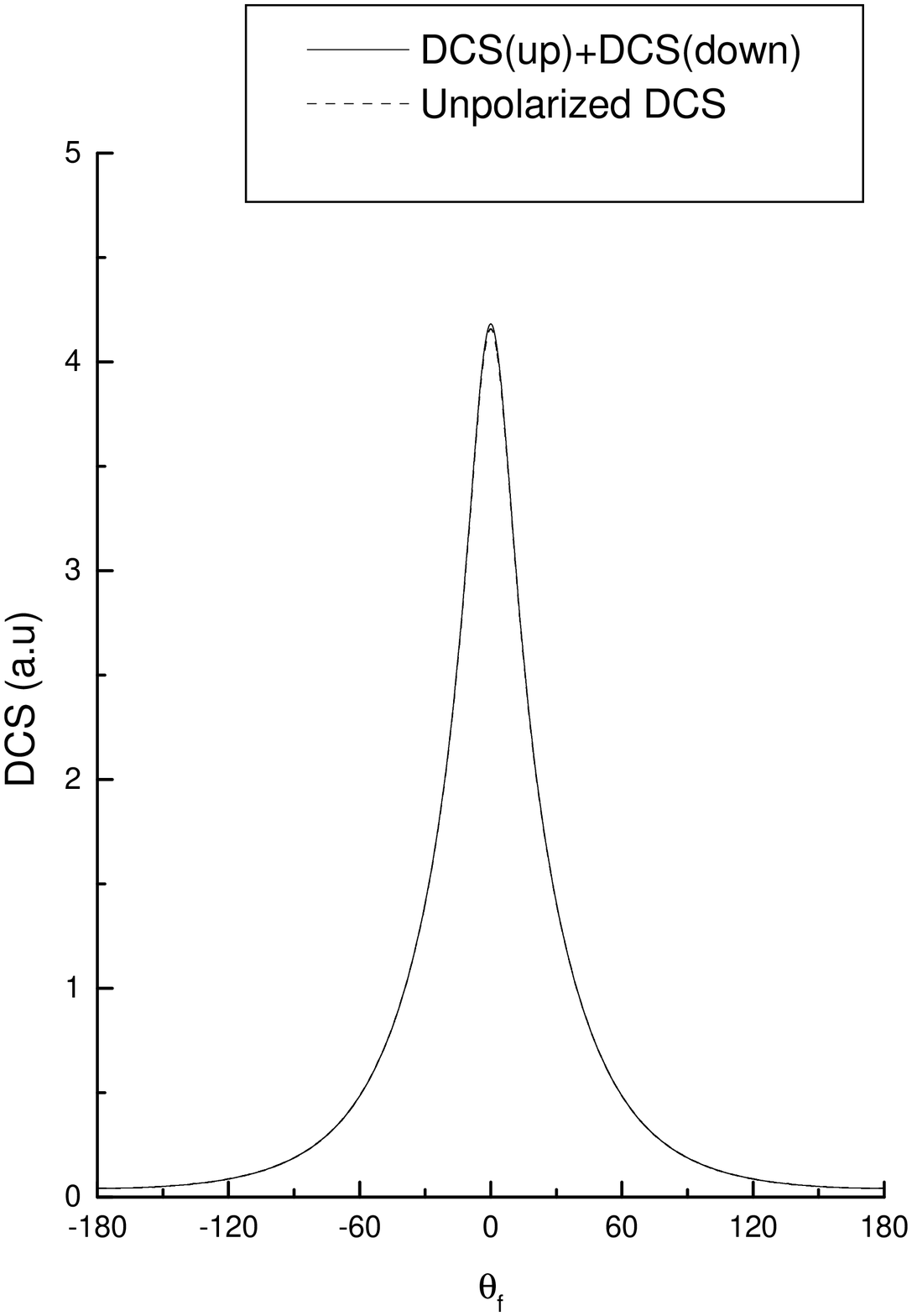}
\caption{First consistency check related to the sum of DCS(up) and
DCS(down) compared to the unpolarized DCS scaled in $10^{-11}$ a.u
as a function of the angle $\theta_{f}$ in degrees for an
electrical field strength of $\mathcal{E}= 1.$ a.u and a
relativistic parameter $\gamma = 2$. The corresponding number of
photons exchanged is $\pm 500$.}
\end{center}
\end{figure}
For this degree of polarization, we obtained a qualitative result
similar to that of a linearly polarized field. We reached the same
conclusion as in \cite{10} that the degree of polarization.
\begin{equation}
P=1-\frac{d\sigma/d\Omega_{f}(\downarrow)}{[d\sigma/d\Omega_{f}(\uparrow)+d\sigma/d\Omega_{f}(\downarrow)]}
\end{equation}
is weakly dependent on the number of photons exchanged. This
degree of polarization $P$ varies as a function of the angle
$\theta_{f}$ for the following geometry ($\theta_{i}=45^{\circ}$,
$\phi_{i}=0$, $\phi_{f}=90^{\circ}$ and $0\leq \theta_{f}\leq
180^{\circ}$). To begin with, we have made simulations concerning
the various DCSs for a set of net number of photons exchanged.
These sets ($\pm 100$, $\pm 200$, $\pm 300$, $\pm 400$, $\pm 500$,
$\pm 1000$) showed that the order of magnitude of the non flip DCS
is close to the unpolarized DCS but the contribution of the flip
DCS is not completely negligible even if it is small compared to
both. The behavior of the three DCSs when the number of photons
exchanged increases has an influence over the numerical values of
the differential cross sections but as we are limited in our
computational capabilities, it is not possible to achieve
numerical convergence but the most important result is that the
sum of DCS(up) and DCS(down) always gives the unpolarized DCS
which is a very important consistency check. Both Figs. (5,6) are accurate illustrations of this consistency check.
\subsection{The non relativistic regime.}
\noindent In this regime, the dressing of the angular coordinates
of $\mathbf{p}_{i}$ as well as those of $\mathbf{p}_{f}$ is not
important. The second and final consistency check is shown in
Fig. (7). First, we give the non relativistic DCS obtained by using
Schr\"{o}dinger-Volkov wave functions :
\begin{equation}
\frac{d\sigma^{(n)}}{d\Omega_{f}}=\frac{4Z^{2}|\mathbf{p}_{f}|}{|\mathbf{p}_{i}|}J^{2}_{n}(\frac{a}{c\omega}\sqrt{\Delta.\hat{x}+\Delta.\hat{y}})\frac{1}{|\mathbf{p}_{i}-\mathbf{p}_{f}+n\mathbf{k}|^{4}}
\end{equation}
The two DCSs are very close, which was to be expected but with
small deviations coming from the unpolarized relativistic DCS in
the limit of small velocities and this is due to the fact that
error propagation is more likely to occur when one is dealing with
very long expressions. This is indeed the case for the latter. We
have summed over $\pm 500$ photons to obtain the curves in
Fig. (7).
\begin{figure}[h]
\begin{center}
\includegraphics[angle=0,width=3. in,height=4. in]{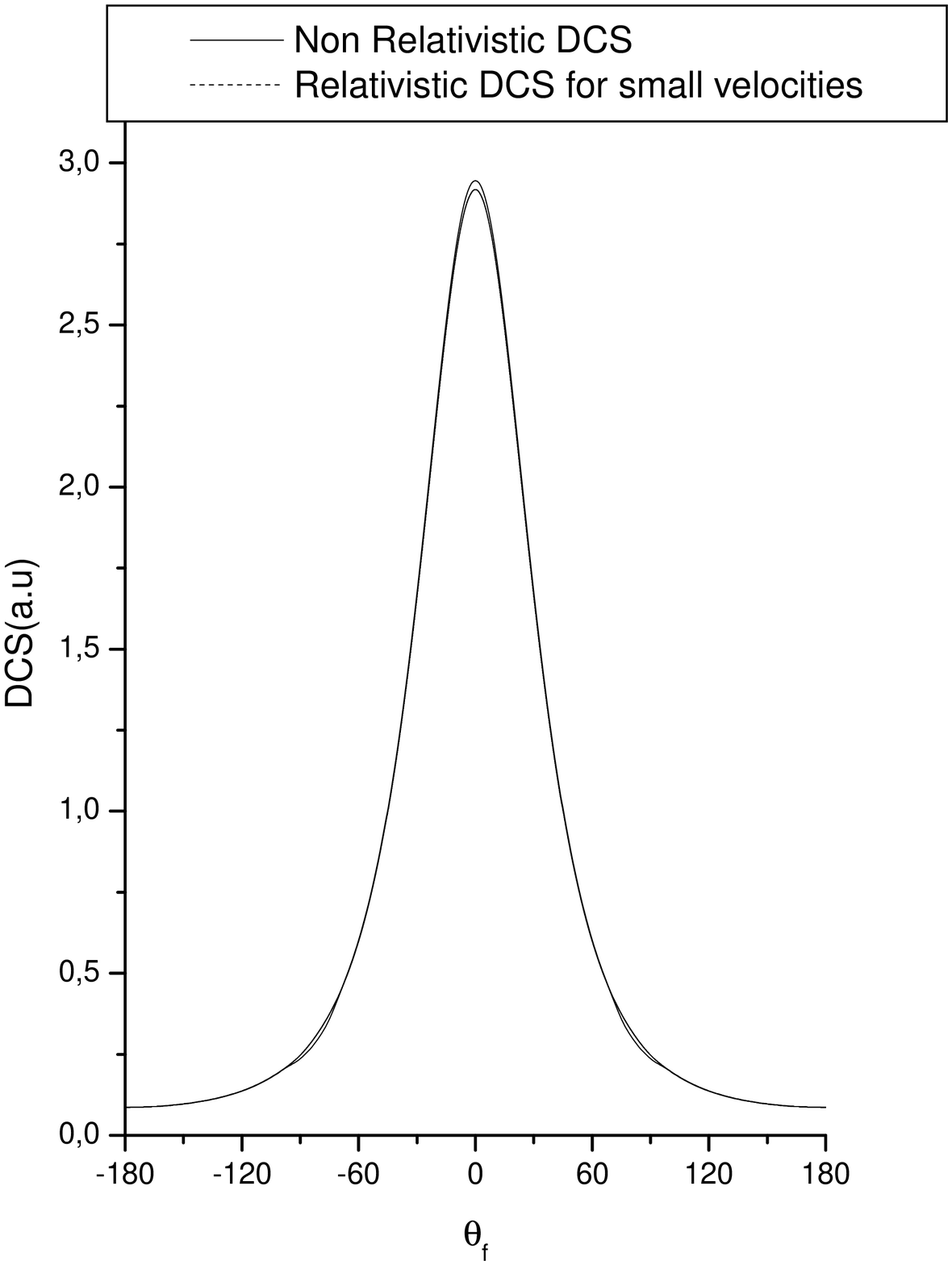}
\caption{Second consistency check in the non relativistic regime
related to the sum of DCS(up) and DCS(down) compared to the non
relativistic DCS scaled in $10^{-4}$ a.u as a function of the
angle $\theta_{f}$ in degrees for an electrical field strength of
$\mathcal{E}= 0.05$ a.u and a relativistic parameter $\gamma =
1.0053$. The corresponding number of photons exchanged is $\pm
500$. For large negative and positive angles, it becomes scaled in
$10^{-5}$ a.u.}
\end{center}
\end{figure}
 There is a very good argument between the two curves and
it is possible by increasing the number of net photons exchanged
to reach the same value as that of the unpolarized DCS in absence
of the laser field. Many similar results were obtained because of
a pseudo sum-rule that was shown by Bunkin and Fedorov as well as
Kroll and Watson \cite{11}.

\begin{figure}[h]
\begin{center}
\includegraphics[angle=0,width=3. in,height=4. in]{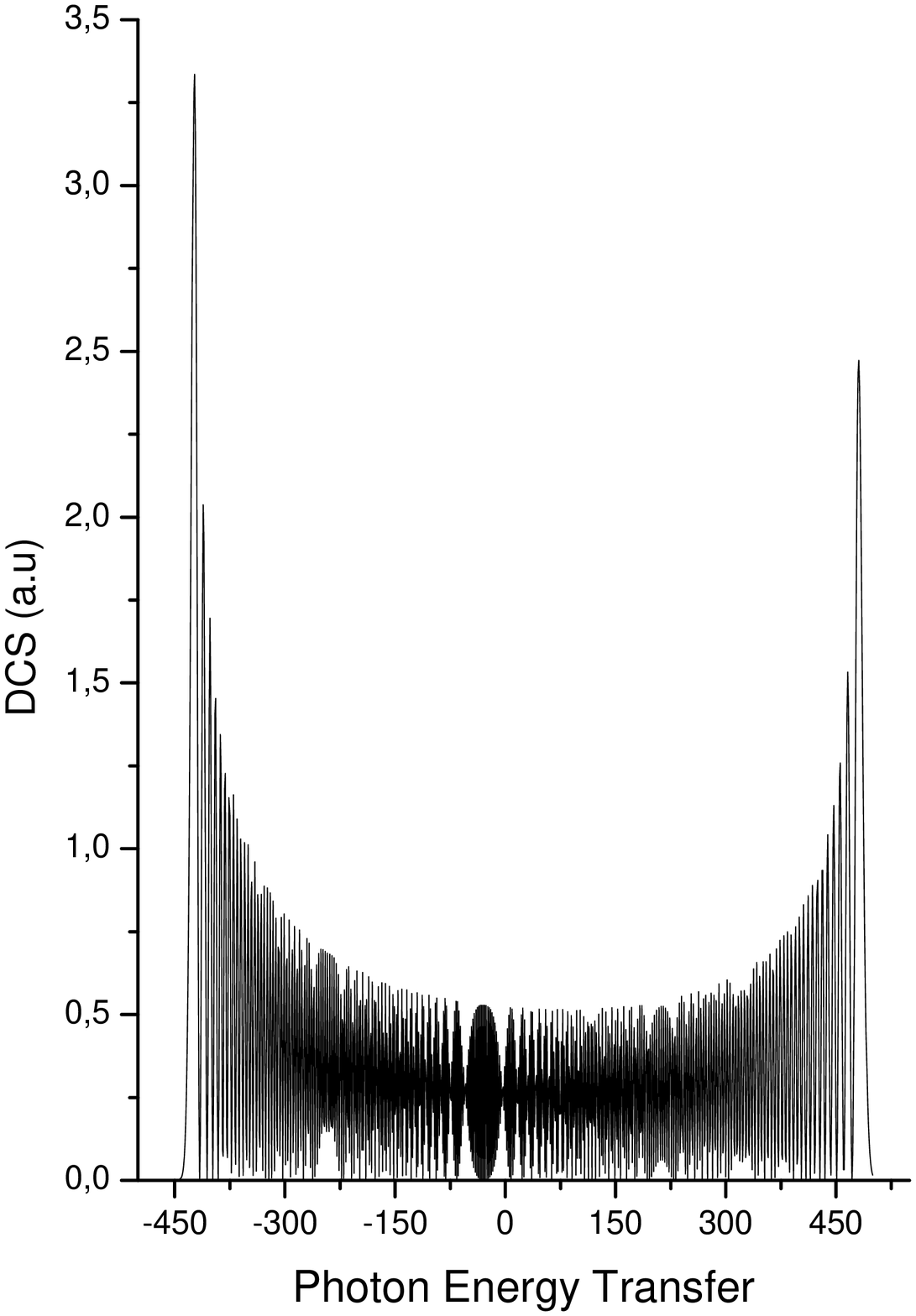}
\caption{Envelope of the two unpolarized DCS in the non
relativistic regime for a number of net photons exchanged $\pm
500$ scaled in units of $10^{-7}$. The two curves are
indistinguishable and are shifted from the elastic peak
corresponding to $\theta_{f}=0^°$. The cutoffs are visible.}
\end{center}
\end{figure}
Even if the non relativistic description does not take into
account the spin of the electron, the agreement between the two
DCSs shows in the one hand, that spin effects are not important
in this regime while on the other hand, given the more complete
and sophisticated relativistic description, the non relativistic
limit of the relativistic DCS always gives results very close to
the non relativistic description. This does not mean that spin
effects are irrelevant but only that their contribution to the
relativistic DCS (for small velocities) is small enough to be
noticeable. As for the cutoffs, one can observe from Fig. (8) that
for the negative part of the spectrum, it is located approximately
at  $-450 $ photons while for the absorptive part of the spectrum,
it is located approximately at  $+470$ photons.
\section{Conclusion.}
\noindent In this work, we have checked that results already known
in the absence of a laser field are also valid in presence of a
laser field, at least in first order of perturbation theory. The
motivation of this study was to compare the order of magnitude for
the various DCSs in the relativistic regime that we already found
in \cite{10} particularly in the case of a laser field with linear
polarization and we have found (see Fig. (6)) that indeed when
scaled in $10^{(-11)}$, both DCSs are of similar order of
magnitude and shape. Moreover, when focusing only on the formalism
for polarized electrons, the main difference between the linear
and the circular polarization of the laser field, is that a
thorough analysis of the contents of the formal expression of
DCS(up) and DCS(down) is more complicated in the case of the
circular polarization than the linear polarization of the laser
field and this is due to the fact that these two DCSs contain non
contracted symbols that we had to deal with for the first time. We
succeeded in the simulations of the main two consistency checks
namely that the sum of DCS(up) and DCS(down) always gives the
unpolarized DCS regardless of the number of net photons exchanged
and also that in the non relativistic limit, both unpolarized DCSs
(non relativistic and relativistic) give very close results. We
also gave the envelopes of the two unpolarized DCSs in both
regimes for the sake of illustration and also to retrieve the
shape of Fig. (3) of \cite{12} which was confirmed since the
scaling is the same ($10^{-7}$) and the asymmetric deviation from
the elastic peak was also found as expected. The envelope for the
relativistic regime was more difficult to obtain since the
stability of the very robust code that evaluates the ordinary
Bessel functions was put on trial because of the very high or very
low arguments of the Bessel functions. Needless to say that
working in simple precision, the results obtained are nonetheless
sound and coherent.
\begin{widetext}
\section{appendix}
\begin{sloppypar}
$ \mathcal{A}(\lambda_i,\lambda_f)=\Big\{2.0 *(k.p_f)^2
*(k.p_i)^2* \lambda_f* \lambda_i *|p_f|^2 *|p_i|^2 *c^8
*\cos(\theta_{if})-2.0 *(k.p_f)^2* (k.p_i)^2 *\lambda_f*
\lambda_i* |p_f|^2 *c^6* \cos(\theta_{if})* E_i^2+2.0 *(k.p_f)^2
*(k.p_i)^2 *\lambda_f* \lambda_i *|p_f|* |p_i| *c^{10}-2.0
*(k.p_f)^2* (k.p_i)^2* \lambda_f* \lambda_i* |p_i|^2 *c^6
*\cos(\theta_{if})* E_f^2+2.0 *(k.p_f)^2* (k.p_i)^2* \lambda_f
*\lambda_i* c^8* \cos(\theta_{if})*E_f* E_i+2.0 (k.p_f)^2
*(k.p_i)^2* \lambda_f* \lambda_i* c^4* \cos(\theta_{if})* E_f^2
*E_i^2* +2.0 *(k.p_f)^2* (k.p_i)^2* |p_f|* |p_i| *c^{10}*
\cos(\theta_{if})+2.0 *(k.p_f)^2* (k.p_i)^2 *c^{12}+2.0 (k.p_f)^2*
(k.p_i)^2* c^8 *E_f *E_i+2.0* (k.p_f)^2 *(k.p_i)* \lambda_f
*\lambda_i* (a^2)* |p_i|^2 *c^4* \cos(\theta_{if})* E_f
*\omega-2.0 *(k.p_f)^2 *(k.p_i) *\lambda_f* \lambda_i*(a^2)* c^2
*\cos(\theta_{if}) *E_f *E_i^2 *\omega-2.0* (k.p_f)^2 *(k.p_i)*
(a^2)* c^6 *E_i *\omega+2.0 *(k.p_f)* (k.p_i)^2* \lambda_f
*\lambda_i* (a^2)* |p_f|^2 *c^4* \cos(\theta_{if})* E_i* \omega
-2.0 *(k.p_f) *(k.p_i)^2 *\lambda_f* \lambda_i* (a^2) *c^2
*\cos(\theta_{if}) *E_f^2 *E_i* \omega-2.0 *(k.p_f) *(k.p_i)^2
*(a^2)* c^6 *E_f *\omega-2.0* (k.p_f) *(k.p_i)* (k.s_f)*
\lambda_f* \lambda_i *(a^2)* |p_f|* |p_i|^2* c^6
*\cos(\theta_{if}) *\omega+2.0 *(k.p_f) *(k.p_i)* (k.s_f)*
\lambda_f *\lambda_i *(a^2)* |p_f|* c^4* \cos(\theta_{if})* E_i^2*
\omega-2.0 *(k.p_f)* (k.p_i) *(k.s_f)* \lambda_f *\lambda_i*
(a^2)* |p_i| *c^8 *\omega-2.0 *(k.p_f)* (k.p_i) *(k.s_i)*
\lambda_f *\lambda_i *(a^2) *|p_f|^2 *|p_i| *c^6*
\cos(\theta_{if}) *\omega-2.0* (k.p_f) *(k.p_i)* (k.s_i)*
\lambda_f* \lambda_i* (a^2)* |p_f|* c^8 *\omega+2.0
*(k.p_f)*(k.p_i) *(k.s_i) *\lambda_f *\lambda_i* (a^2) *|p_i| *c^4
*\cos(\theta_{if})* E_f^2 *\omega-(k.p_f) *(k.p_i) *\lambda_f*
\lambda_i*(a^2)^2 *|p_f|* |p_i|* c^2 *\omega^2+(k.p_f)*(k.p_i)*
\lambda_f* \lambda_i *(a^2)^2*\cos(\theta_{if})* E_f* E_i*
\omega^2+2.0 *(k.p_f)* (k.p_i) *\lambda_f* \lambda_i *(a^2)
*|p_f|^2* |p_i|^2* c^4 *\cos(\theta_{if}) *\omega^2-2.0* (k.p_f)
(k.p_i) *\lambda_f *\lambda_i *(a^2) *|p_f|^2 *c^2
*\cos(\theta_{if}) *E_i^2 *\omega^2 +2.0 *(k.p_f) (k.p_i)*
\lambda_f *\lambda_i *(a^2)* |p_f|* |p_i| *c^6 *\omega^2 -2.0
*(k.p_f) *(k.p_i)* \lambda_f *\lambda_i *(a^2)* |p_i|^2 *c^2*
\cos(\theta_{if})* E_f^2 *\omega^2-2.0 *(k.p_f)* (k.p_i)* \lambda_f
*\lambda_i *(a^2)* c^4* \cos(\theta_{if})* E_f *E_i* \omega^2+2.0
*(k.p_f) (k.p_i) \lambda_f *\lambda_i *(a^2) *\cos(\theta_{if})*
E_f^2 *E_i^2* \omega^2 +(k.p_f) (k.p_i) *(a^2)^2* c^4
*\omega^2-2.0 *(k.p_f)* (k.p_i)* (a^2)* |p_f| *|p_i|* c^6*
\cos(\theta_{if})* \omega^2-2.0 (k.p_f) (k.p_i) *(a^2)* c^8
*\omega^2 +2.0 *(k.p_f)* (k.p_i) *(a^2) c^4 *E_f* E_i*
\omega^2+(k.p_f) *(k.s_i) *\lambda_f *\lambda_i* (a^2)^2 *|p_f|*
c^2 *E_i *\omega^2-(k.p_f) (k.s_i)* \lambda_f* \lambda_i *(a^2)^2
*|p_i|* c^2 *\cos(\theta_{if})* E_f* \omega^2-(k.p_i) *(k.s_f)
*\lambda_f* \lambda_i *(a^2)^2 *|p_f|* c^2 *\cos(\theta_{if})* E_i
*\omega^2+(k.p_i)* (k.s_f) *\lambda_f* \lambda_i* (a^2)^2* |p_i|*
c^2 *E_f *\omega^2 + (k.s_f)* (k.s_i) \lambda_f *\lambda_i*
(a^2)^2* |p_f|* |p_i|* c^4 *\cos(\theta_{if}) *\omega^2+(k.s_f)
(k.s_i)* \lambda_f *\lambda_i* (a^2)^2* c^6* \omega^2-(k.s_f)
(k.s_i) *\lambda_f* \lambda_i *(a^2)^2*c^2* E_f* E_i
*\omega^2\Big\}/[2.0 *(k.p_f)^2 *(k.p_i)^2* c^8] $
\end{sloppypar}

\begin{sloppypar}
$\mathcal{B}(\lambda_i,\lambda_f)=\Big\{(a_1.p_f)*(a_1.p_i)*(k.p_f)^2*(k.s_f)*(k.s_i)*\lambda_f*\lambda_i*c^8-2.0*(a_1.p_f)*(a_1.p_i)*
    (k.p_f)^2*(k.s_i)*\lambda_f*\lambda_i*|p_f|*c^6*\omega-2.0*(a_1.p_f)*(a_1.p_i)*(k.p_f)*(k.p_i)*
     (k.s_f)*(k.s_i)*\lambda_f*\lambda_i*c^8+2.0*(a_1.p_f)*(a_1.p_i)*(k.p_f)*(k.p_i)*(k.s_f)*\lambda_f*\lambda_i*
     |p_i|*c^6*\omega+2.0*(a_1.p_f)*(a_1.p_i)*(k.p_f)*(k.p_i)*(k.s_i)*\lambda_f*\lambda_i*|p_f|*c^6*
     \omega-2.0*(a_1.p_f)*(a_1.p_i)*(k.p_f)*(k.p_i)*\lambda_f*\lambda_i*|p_f|*|p_i|*c^4*\omega^2+2.0*
     (a_1.p_f)*(a_1.p_i)*(k.p_f)*(k.p_i)*\lambda_f*\lambda_i*c^2*\cos(\theta_{if})*E_f*E_i*\omega^2+2.0*
     (a_1.p_f)*(a_1.p_i)*(k.p_f)*(k.p_i)*c^6*\omega^2+(a_1.p_f)*(a_1.p_i)*(k.p_i)^2*(k.s_f)*(k.s_i)*
     \lambda_f*\lambda_i*c^8-2.0*(a_1.p_f)*(a_1.p_i)*(k.p_i)^2*(k.s_f)*\lambda_f*\lambda_i*|p_i|*c^6*\omega-
      (a_1.p_f)*(a_1.s_i)*(k.p_f)^2*(k.p_i)*(k.s_f)*\lambda_f*\lambda_i*c^8+2.0*(a_1.p_f)*(a_1.s_i)*(k.p_f)^2
      *(k.p_i)*\lambda_f*\lambda_i*|p_f|*c^6*\omega+2.0*(a_1.p_f)*(a_1.s_i)*(k.p_f)*(k.p_i)^2*(k.s_f)*
      \lambda_f*\lambda_i*c^8-2.0*(a_1.p_f)*(a_1.s_i)*(k.p_f)*(k.p_i)^2*\lambda_f*\lambda_i*|p_f|*c^6*\omega-
     2.0*(a_1.p_f)*(a_1.s_i)*(k.p_f)*(k.p_i)*(k.s_f)*\lambda_f*\lambda_i*c^6*E_i*\omega+2.0*(a_1.p_f)*
     (a_1.s_i)*(k.p_f)*(k.p_i)*\lambda_f*\lambda_i*|p_f|*c^4*E_i*\omega^2-2.0*(a_1.p_f)*(a_1.s_i)*
      (k.p_f)*(k.p_i)*\lambda_f*\lambda_i*|p_i|*c^4*\cos(\theta_{if})*E_f*\omega^2-(a_1.p_f)*(a_1.s_i)*(k.p_i)^3
    *(k.s_f)*\lambda_f*\lambda_i*c^8+2.0*(a_1.p_f)*(a_1.s_i)*(k.p_i)^2*(k.s_f)*\lambda_f*\lambda_i*c^6*
     E_i*\omega-(a_1.p_i)*(a_1.s_f)*(k.p_f)^3*(k.s_i)*\lambda_f*\lambda_i*c^8+2.0*(a_1.p_i)*(a_1.s_f)*
     (k.p_f)^2*(k.p_i)*(k.s_i)*\lambda_f*\lambda_i*c^8-2.0*(a_1.p_i)*(a_1.s_f)*(k.p_f)^2*(k.p_i)*\lambda_f*
    \lambda_i*|p_i|*c^6*\omega+2.0*(a_1.p_i)*(a_1.s_f)*(k.p_f)^2*(k.s_i)*\lambda_f*\lambda_i*c^6*E_f*
     \omega-(a_1.p_i)*(a_1.s_f)*(k.p_f)*(k.p_i)^2*(k.s_i)*\lambda_f*\lambda_i*c^8+2.0*(a_1.p_i)*(a_1.s_f)*(k.p_f)*(k.p_i)^2*\lambda_f*\lambda_i*|p_i|*c^6*\omega-2.0*(a_1.p_i)*
     (a_1.s_f)*(k.p_f)*(k.p_i)*(k.s_i)*\lambda_f*\lambda_i*c^6*E_f*\omega-2.0*(a_1.p_i)*(a_1.s_f)*(k.p_f)*
     (k.p_i)*\lambda_f*\lambda_i*|p_f|*c^4*\cos(\theta_{if})*E_i*\omega^2+2.0*(a_1.p_i)*(a_1.s_f)*(k.p_f)*
     (k.p_i)*\lambda_f*\lambda_i*|p_i|*c^4*E_f*\omega^2+(a_1.s_f)*(a_1.s_i)*(k.p_f)^3*(k.p_i)*\lambda_f*
     \lambda_i*c^8-2.0*(a_1.s_f)*(a_1.s_i)*(k.p_f)^2*(k.p_i)^2*\lambda_f*\lambda_i*c^8-2.0*(a_1.s_f)*
    (a_1.s_i)*(k.p_f)^2*(k.p_i)*\lambda_f*\lambda_i*c^6*E_f*\omega+2.0*(a_1.s_f)*(a_1.s_i)*(k.p_f)^2*
     (k.p_i)*\lambda_f*\lambda_i*c^6*E_i*\omega+(a_1.s_f)*(a_1.s_i)*(k.p_f)*(k.p_i)^3*\lambda_f*\lambda_i*c^
    8+2.0*(a_1.s_f)*(a_1.s_i)*(k.p_f)*(k.p_i)^2*\lambda_f*\lambda_i*c^6*E_f*\omega-2.0*(a_1.s_f)*
    (a_1.s_i)*(k.p_f)*(k.p_i)^2*\lambda_f*\lambda_i*c^6*E_i*\omega+2.0*(a_1.s_f)*(a_1.s_i)*(k.p_f)*
    (k.p_i)*\lambda_f*\lambda_i*|p_f|*|p_i|*c^6*\cos(\theta_{if})*\omega^2+2.0*(a_1.s_f)*(a_1.s_i)*(k.p_f)*
    (k.p_i)*\lambda_f*\lambda_i*c^8*\omega^2-2.0*(a_1.s_f)*(a_1.s_i)*(k.p_f)*(k.p_i)*\lambda_f*\lambda_i*c^4
    *E_f*E_i*\omega^2+(a_2.p_f)*(a_2.p_i)*(k.p_f)^2*(k.s_f)*(k.s_i)*\lambda_f*\lambda_i*c^8-2.0
    *(a_2.p_f)*(a_2.p_i)*(k.p_f)^2*(k.s_i)*\lambda_f*\lambda_i*|p_f|*c^6*\omega-2.0*(a_2.p_f)*(a_2.p_i)*
     (k.p_f)*(k.p_i)*(k.s_f)*(k.s_i)*\lambda_f*\lambda_i*c^8+2.0*(a_2.p_f)*(a_2.p_i)*(k.p_f)*(k.p_i)*(k.s_f)*
    \lambda_f*\lambda_i*|p_i|*c^6*\omega+2.0*(a_2.p_f)*(a_2.p_i)*(k.p_f)*(k.p_i)*(k.s_i)*\lambda_f*\lambda_i*|p_f|
     *c^6*\omega-2.0*(a_2.p_f)*(a_2.p_i)*(k.p_f)*(k.p_i)*\lambda_f*\lambda_i*|p_f|*|p_i|*c^4*\omega^
      2+2.0*(a_2.p_f)*(a_2.p_i)*(k.p_f)*(k.p_i)*\lambda_f*\lambda_i*c^2*\cos(\theta_{if})*E_f*E_i*\omega^2+
    2.0*(a_2.p_f)*(a_2.p_i)*(k.p_f)*(k.p_i)*c^6*\omega^2+(a_2.p_f)*(a_2.p_i)*(k.p_i)^2*(k.s_f)*
      (k.s_i)*\lambda_f*\lambda_i*c^8-2.0*(a_2.p_f)*(a_2.p_i)*(k.p_i)^2*(k.s_f)*\lambda_f*\lambda_i*|p_i|*c^6*\omega-(a_2.p_f)*
      (a_2.s_i)*(k.p_f)^2*(k.p_i)*(k.s_f)*\lambda_f*\lambda_i*c^8+2.0*(a_2.p_f)*(a_2.s_i)*(k.p_f)^2*(k.p_i)
      *\lambda_f*\lambda_i*|p_f|*c^6*\omega+2.0*(a_2.p_f)*(a_2.s_i)*(k.p_f)*(k.p_i)^2*(k.s_f)*\lambda_f*\lambda_i
      *c^8-2.0*(a_2.p_f)*(a_2.s_i)*(k.p_f)*(k.p_i)^2*\lambda_f*\lambda_i*|p_f|*c^6*\omega-2.0*
      (a_2.p_f)*(a_2.s_i)*(k.p_f)*(k.p_i)*(k.s_f)*\lambda_f*\lambda_i*c^6*E_i*\omega+2.0*(a_2.p_f)*(a_2.s_i)*
      (k.p_f)*(k.p_i)*\lambda_f*\lambda_i*|p_f|*c^4*E_i*\omega^2-2.0*(a_2.p_f)*(a_2.s_i)*(k.p_f)*
      (k.p_i)*\lambda_f*\lambda_i*|p_i|*c^4*\cos(\theta_{if})*E_f*\omega^2-(a_2.p_f)*(a_2.s_i)*(k.p_i)^3*(k.s_f)
      *\lambda_f*\lambda_i*c^8+2.0*(a_2.p_f)*(a_2.s_i)*(k.p_i)^2*(k.s_f)*\lambda_f*\lambda_i*c^6*E_i*
      \omega-(a_2.p_i)*(a_2.s_f)*(k.p_f)^3*(k.s_i)*\lambda_f*\lambda_i*c^8+2.0*(a_2.p_i)*(a_2.s_f)*(k.p_f)^2
      *(k.p_i)*(k.s_i)*\lambda_f*\lambda_i*c^8-2.0*(a_2.p_i)*(a_2.s_f)*(k.p_f)^2*(k.p_i)*\lambda_f*\lambda_i*|p_i|
      *c^6*\omega+2.0*(a_2.p_i)*(a_2.s_f)*(k.p_f)^2*(k.s_i)*\lambda_f*\lambda_i*c^6*E_f*\omega-
      (a_2.p_i)*(a_2.s_f)*(k.p_f)*(k.p_i)^2*(k.s_i)*\lambda_f*\lambda_i*c^8+2.0*(a_2.p_i)*(a_2.s_f)*(k.p_f)*
      (k.p_i)^2*\lambda_f*\lambda_i*|p_i|*c^6*\omega-2.0*(a_2.p_i)*(a_2.s_f)*(k.p_f)*(k.p_i)*(k.s_i)*\lambda_f
      *\lambda_i*c^6*E_f*\omega-2.0*(a_2.p_i)*(a_2.s_f)*(k.p_f)*(k.p_i)*\lambda_f*\lambda_i*|p_f|*c^4*
      \cos(\theta_{if})*E_i*\omega^2+2.0*(a_2.p_i)*(a_2.s_f)*(k.p_f)*(k.p_i)*\lambda_f*\lambda_i*|p_i|*c^4*
     E_f*\omega^2+(a_2.s_f)*(a_2.s_i)*(k.p_f)^3*(k.p_i)*\lambda_f*\lambda_i*c^8-2.0*(a_2.s_f)*(a_2.s_i)*
      (k.p_f)^2*(k.p_i)^2*\lambda_f*\lambda_i*c^8-2.0*(a_2.s_f)*(a_2.s_i)*(k.p_f)^2*(k.p_i)*\lambda_f*
     \lambda_i*c^6*E_f*\omega+2.0*(a_2.s_f)*(a_2.s_i)*(k.p_f)^2*(k.p_i)*\lambda_f*\lambda_i*c^6*E_i
      *\omega+(a_2.s_f)*(a_2.s_i)*(k.p_f)*(k.p_i)^3*\lambda_f*\lambda_i*c^8+2.0*(a_2.s_f)*(a_2.s_i)*(k.p_f)*
      (k.p_i)^2*\lambda_f*\lambda_i*c^6*E_f*\omega-2.0*(a_2.s_f)*(a_2.s_i)*(k.p_f)*(k.p_i)^2*\lambda_f*
      \lambda_i*c^6*E_i*\omega+2.0*(a_2.s_f)*(a_2.s_i)*(k.p_f)*(k.p_i)*\lambda_f*\lambda_i*|p_f|*|p_i|*c^
      6*\cos(\theta_{if})*\omega^2+2.0*(a_2.s_f)*(a_2.s_i)*(k.p_f)*(k.p_i)*\lambda_f*\lambda_i*c^8*\omega^2-
      2.0*(a_2.s_f)*(a_2.s_i)*(k.p_f)*(k.p_i)*\lambda_f*\lambda_i*c^4*E_f*E_i*\omega^2+(k.p_f)^3*
      (k.p_i)*\lambda_f*\lambda_i*(a^2)*|p_f|*|p_i|*c^6+(k.p_f)^3*(k.p_i)*\lambda_f*\lambda_i*(a^2)*c^4*
      \cos(\theta_{if})*E_f*E_i+(k.p_f)^3*(k.p_i)*(a^2)*c^8-(k.p_f)^3*(k.s_i)*\lambda_f*\lambda_i*(a^2)*
      |p_f|*c^6*E_i-(k.p_f)^3*(k.s_i)*\lambda_f*\lambda_i*(a^2)*|p_i|*c^6*\cos(\theta_{if})*E_f-2.0*
      (k.p_f)^2*(k.p_i)^2*\lambda_f*\lambda_i*(a^2)*|p_f|*|p_i|*c^6-2.0*(k.p_f)^2*(k.p_i)^2*
      \lambda_f*\lambda_i*(a^2)*c^4*\cos(\theta_{if})*E_f*E_i-2.0*(k.p_f)^2*(k.p_i)^2*(a^2)*c^8-
      (k.p_f)^2*(k.p_i)*(k.s_f)*\lambda_f*\lambda_i*(a^2)*|p_f|*c^6*\cos(\theta_{if})*E_i-(k.p_f)^2*(k.p_i)
      *(k.s_f)*\lambda_f*\lambda_i*(a^2)*|p_i|*c^6*E_f+2.0*(k.p_f)^2*(k.p_i)*(k.s_i)*\lambda_f*\lambda_i*
      (a^2)*|p_f|*c^6*E_i+2.0*(k.p_f)^2*(k.p_i)*(k.s_i)*\lambda_f*\lambda_i*(a^2)*|p_i|*c^6*
      \cos(\theta_{if})*E_f+2.0*(k.p_f)^2*(k.p_i)*\lambda_f*\lambda_i*(a^2)*|p_f|^2*c^4*\cos(\theta_{if})*E_i*
      \omega-2.0*(k.p_f)^2*(k.p_i)*\lambda_f*\lambda_i*(a^2)*|p_i|^2*c^4*\cos(\theta_{if})*E_f*\omega-
      2.0*(k.p_f)^2*(k.p_i)*\lambda_f*\lambda_i*(a^2)*c^2*\cos(\theta_{if})*E_f^2*E_i*\omega+2.0*
      (k.p_f)^2*(k.p_i)*\lambda_f*\lambda_i*(a^2)*c^2*\cos(\theta_{if})*E_f*E_i^2*\omega-(k.p_f)^2*
      (k.p_i)*\lambda_f*(a^2)*|p_f|*c^7*\omega-(k.p_f)^2*(k.p_i)*\lambda_f*(a^2)*c^6*\cos(\theta_f)*E_f*\omega+(k.p_f)^2*(k.p_i)*\lambda_i
      *(a^2)*|p_i|*c^7*\omega-(k.p_f)^2*(k.p_i)*\lambda_i*(a^2)*c^6*\cos(\theta_i)*E_i*\omega-
      2.0*(k.p_f)^2*(k.p_i)*(a^2)*c^6*E_f*\omega+2.0*(k.p_f)^2*(k.p_i)*(a^2)*c^6*
      E_i*\omega+(k.p_f)^2*(k.s_f)*(k.s_i)*\lambda_f*\lambda_i*(a^2)*|p_f|*|p_i|*c^8*\cos(\theta_{if})+
      (k.p_f)^2*(k.s_f)*(k.s_i)*\lambda_f*\lambda_i*(a^2)*c^{10}+(k.p_f)^2*(k.s_f)*(k.s_i)*\lambda_f*\lambda_i*
      (a^2)*c^6*E_f*E_i-2.0*(k.p_f)^2*(k.s_i)*\lambda_f*\lambda_i*(a^2)*|p_f|^2*|p_i|*c^6
      *\cos(\theta_{if})*\omega-2.0*(k.p_f)^2*(k.s_i)*\lambda_f*\lambda_i*(a^2)*|p_f|*c^8*\omega+2.0*
      (k.p_f)^2*(k.s_i)*\lambda_f*\lambda_i*(a^2)*|p_i|*c^4*\cos(\theta_{if})*E_f^2*\omega-(k.p_f)^2*
      (k.s_i)*\lambda_i*(a^2)*|p_f|*c^8*\cos(\theta_f)*\omega+(k.p_f)^2*(k.s_i)*\lambda_i*(a^2)*|p_i|*c^8
      *\cos(\theta_i)*\omega-(k.p_f)^2*(k.s_i)*\lambda_i*(a^2)*c^7*E_f*\omega-(k.p_f)^2*(k.s_i)*
      \lambda_i*(a^2)*c^7*E_i*\omega+(k.p_f)*(k.p_i)^3*\lambda_f*\lambda_i*(a^2)*|p_f|*|p_i|*c^6+
      (k.p_f)*(k.p_i)^3*\lambda_f*\lambda_i*(a^2)*c^4*\cos(\theta_{if})*E_f*E_i+(k.p_f)*(k.p_i)^3*(a^2)*
      c^8+2.0*(k.p_f)*(k.p_i)^2*(k.s_f)*\lambda_f*\lambda_i*(a^2)*|p_f|*c^6*\cos(\theta_{if})*E_i+2.0
      *(k.p_f)*(k.p_i)^2*(k.s_f)*\lambda_f*\lambda_i*(a^2)*|p_i|*c^6*E_f-(k.p_f)*(k.p_i)^2*(k.s_i)*
      \lambda_f*\lambda_i*(a^2)*|p_f|*c^6*E_i-(k.p_f)*(k.p_i)^2*(k.s_i)*\lambda_f*\lambda_i*(a^2)*|p_i|*c
      ^6*\cos(\theta_{if})*E_f-2.0*(k.p_f)*(k.p_i)^2*\lambda_f*\lambda_i*(a^2)*|p_f|^2*c^4*\cos(\theta_{if})*
      E_i*\omega+2.0*(k.p_f)*(k.p_i)^2*\lambda_f*\lambda_i*(a^2)*|p_i|^2*c^4*\cos(\theta_{if})*E_f*
       \omega+2.0*(k.p_f)*(k.p_i)^2*\lambda_f*\lambda_i*(a^2)*c^2*\cos(\theta_{if})*E_f^2*E_i*\omega-
      2.0*(k.p_f)*(k.p_i)^2*\lambda_f*\lambda_i*(a^2)*c^2*\cos(\theta_{if})*E_f*E_i^2*\omega+2.0*
      (k.p_f)*(k.p_i)^2*\lambda_f*(a^2)*|p_f|*c^7*\omega-2.0*(k.p_f)*(k.p_i)^2*\lambda_i*(a^2)*
      |p_i|*c^7*\omega+2.0*(k.p_f)*(k.p_i)^2*(a^2)*c^6*E_f*\omega-2.0*(k.p_f)*(k.p_i)
      ^2*(a^2)*c^6*E_i*\omega-2.0*(k.p_f)*(k.p_i)*(k.s_f)*(k.s_i)*\lambda_f*\lambda_i*(a^2)*|p_f|
      *|p_i|*c^8*\cos(\theta_{if})-2.0*(k.p_f)*(k.p_i)*(k.s_f)*(k.s_i)*\lambda_f*\lambda_i*(a^2)*c^{10}-2.0*
      (k.p_f)*(k.p_i)*(k.s_f)*(k.s_i)*\lambda_f*\lambda_i*(a^2)*c^6*E_f*E_i+2.0*(k.p_f)*(k.p_i)*
      (k.s_f)*\lambda_f*\lambda_i*(a^2)*|p_f|*|p_i|^2*c^6*\cos(\theta_{if})*\omega-2.0*(k.p_f)*(k.p_i)*(k.s_f)
      *\lambda_f*\lambda_i*(a^2)*|p_f|*c^4*\cos(\theta_{if})*E_i^2*\omega+2.0*(k.p_f)*(k.p_i)*(k.s_f)*
      \lambda_f*\lambda_i*(a^2)*|p_i|*c^8*\omega+2.0*(k.p_f)*(k.p_i)*(k.s_i)*\lambda_f*\lambda_i*(a^2)*|p_f|^
      2*|p_i|*c^6*\cos(\theta_{if})*\omega+2.0*(k.p_f)*(k.p_i)*(k.s_i)*\lambda_f*\lambda_i*(a^2)*|p_f|*c^8*
      \omega-2.0*(k.p_f)*(k.p_i)*(k.s_i)*\lambda_f*\lambda_i*(a^2)*|p_i|*c^4*\cos(\theta_{if})*E_f^2*\omega
      +2.0*(k.p_f)*(k.p_i)*(k.s_i)*\lambda_i*(a^2)*c^7*E_f*\omega+2.0*(k.p_f)*(k.p_i)*(k.s_i)*
      \lambda_i*(a^2)*c^7*E_i*\omega-2.0*(k.p_f)*(k.p_i)*\lambda_f*\lambda_i*(a^2)*|p_f|^2*|p_i|^2
      *c^4*\cos(\theta_{if})*\omega^2+2.0*(k.p_f)*(k.p_i)*\lambda_f*\lambda_i*(a^2)*|p_f|^2*c^2*
      \cos(\theta_{if})*E_i^2*\omega^2-2.0*(k.p_f)*(k.p_i)*\lambda_f*\lambda_i*(a^2)*|p_f|*|p_i|*c^6*
      \omega^2+2.0*(k.p_f)*(k.p_i)*\lambda_f*\lambda_i*(a^2)*|p_i|^2*c^2*\cos(\theta_{if})*E_f^2*
      \omega^2+2.0*(k.p_f)*(k.p_i)*\lambda_f*\lambda_i*(a^2)*c^4*\cos(\theta_{if})*E_f*E_i*\omega^2-2.0*(k.p_f)*(k.p_i)*\lambda_f*\lambda_i*(a^2)*\cos(\theta_{if})*E_f^2*E_i^2*\omega^2-
      2.0*(k.p_f)*(k.p_i)*\lambda_f*(a^2)*|p_f|^2*c^6*\cos(\theta_f)*\omega^2+2.0*(k.p_f)*(k.p_i)*
      \lambda_f*(a^2)*|p_f|*|p_i|*c^6*\cos(\theta_i)*\omega^2-2.0*(k.p_f)*(k.p_i)*\lambda_f*(a^2)*|p_f|*c
      ^5*E_i*\omega^2+2.0*(k.p_f)*(k.p_i)*\lambda_f*(a^2)*c^4*\cos(\theta_f)*E_f^2*\omega
      ^2-2.0*(k.p_f)*(k.p_i)*\lambda_i*(a^2)*|p_i|*c^5*E_f*\omega^2+2.0*(k.p_f)*(k.p_i)*
      \lambda_i*(a^2)*c^4*\cos(\theta_i)*E_f*E_i*\omega^2+2.0*(k.p_f)*(k.p_i)*(a^2)*|p_f|*|p_i|*
      c^6*\cos(\theta_{if})*\omega^2+2.0*(k.p_f)*(k.p_i)*(a^2)*c^8*\omega^2-2.0*(k.p_f)*
      (k.p_i)*(a^2)*c^4*E_f*E_i*\omega^2-(k.p_i)^3*(k.s_f)*\lambda_f*\lambda_i*(a^2)*|p_f|*c
      ^6*\cos(\theta_{if})*E_i-(k.p_i)^3*(k.s_f)*\lambda_f*\lambda_i*(a^2)*|p_i|*c^6*E_f-(k.p_i)^3*
      \lambda_f*(a^2)*|p_f|*c^7*\omega+(k.p_i)^3*\lambda_f*(a^2)*c^6*\cos(\theta_f)*E_f*\omega+(k.p_i)
      ^3*\lambda_i*(a^2)*|p_i|*c^7*\omega+(k.p_i)^3*\lambda_i*(a^2)*c^6*\cos(\theta_i)*E_i*\omega
      +(k.p_i)^2*(k.s_f)*(k.s_i)*\lambda_f*\lambda_i*(a^2)*|p_f|*|p_i|*c^8*\cos(\theta_{if})+(k.p_i)^2*(k.s_f)
      *(k.s_i)*\lambda_f*\lambda_i*(a^2)*c^{10}+(k.p_i)^2*(k.s_f)*(k.s_i)*\lambda_f*\lambda_i*(a^2)*c^6*E_f
      *E_i-2.0*(k.p_i)^2*(k.s_f)*\lambda_f*\lambda_i*(a^2)*|p_f|*|p_i|^2*c^6*\cos(\theta_{if})*\omega+
      2.0*(k.p_i)^2*(k.s_f)*\lambda_f*\lambda_i*(a^2)*|p_f|*c^4*\cos(\theta_{if})*E_i^2*\omega-2.0*
      (k.p_i)^2*(k.s_f)*\lambda_f*\lambda_i*(a^2)*|p_i|*c^8*\omega+(k.p_i)^2*(k.s_i)*\lambda_i*(a^2)*
      |p_f|*c^8*\cos(\theta_f)*\omega-(k.p_i)^2*(k.s_i)*\lambda_i*(a^2)*|p_i|*c^8*\cos(\theta_i)*\omega-
      (k.p_i)^2*(k.s_i)*\lambda_i*(a^2)*c^7*E_f*\omega-(k.p_i)^2*(k.s_i)*\lambda_i*(a^2)*c^7*E_i*\omega+2.0*(k.p_i)^2*\lambda_f*(a^2)*
      |p_f|*c^5*E_i*\omega^2-2.0*(k.p_i)^2*\lambda_f*(a^2)*c^4*\cos(\theta_f)*E_f*E_i*
      \omega^2-2.0*(k.p_i)^2*\lambda_i*(a^2)*|p_f|*|p_i|*c^6*\cos(\theta_f)*\omega^2+2.0*
      (k.p_i)^2*\lambda_i*(a^2)*|p_i|^2*c^6*\cos(\theta_i)*\omega^2+2.0*(k.p_i)^2*\lambda_i*(a^2)*
      |p_i|*c^5*E_f*\omega^2-2.0*(k.p_i)^2*\lambda_i*(a^2)*c^4*\cos(\theta_i)*E_i^2*
      \omega^2\Big\}
      /[4.0*(k.p_f)^2*(k.p_i)^2*c^8]
$
\end{sloppypar}

\begin{sloppypar}
$  \mathcal{C}(\lambda_i,\lambda_f)=\Big\{\cos(2
\phi_0)*(a_1.p_f)*(a_1.p_i)*(k.p_f)^2*(k.s_f)*(k.s_i)*\lambda_f*\lambda_i*c^6-2.0
      *\cos(2 \phi_0)*(a_1.p_f)*(a_1.p_i)*(k.p_f)^2*(k.s_i)*\lambda_f*\lambda_i*|p_f|*c^4*\omega-
      2.0*\cos(2 \phi_0)*(a_1.p_f)*(a_1.p_i)*(k.p_f)*(k.p_i)*(k.s_f)*(k.s_i)*\lambda_f*\lambda_i*c^6+
      2.0*\cos(2 \phi_0)*(a_1.p_f)*(a_1.p_i)*(k.p_f)*(k.p_i)*(k.s_f)*\lambda_f*\lambda_i*|p_i|*c^4*
      \omega+2.0*\cos(2 \phi_0)*(a_1.p_f)*(a_1.p_i)*(k.p_f)*(k.p_i)*(k.s_i)*\lambda_f*\lambda_i*|p_f|*c
      ^4*\omega-2.0*\cos(2 \phi_0)*(a_1.p_f)*(a_1.p_i)*(k.p_f)*(k.p_i)*\lambda_f*\lambda_i*|p_f|*|p_i|
      *c^2*\omega^2+2.0*\cos(2 \phi_0)*(a_1.p_f)*(a_1.p_i)*(k.p_f)*(k.p_i)*\lambda_f*\lambda_i*
      \cos(\theta_{if})*E_f*E_i*\omega^2+2.0*\cos(2 \phi_0)*(a_1.p_f)*(a_1.p_i)*(k.p_f)*(k.p_i)*c
      ^4*\omega^2+\cos(2 \phi_0)*(a_1.p_f)*(a_1.p_i)*(k.p_i)^2*(k.s_f)*(k.s_i)*\lambda_f*\lambda_i*
      c^6-2.0*\cos(2 \phi_0)*(a_1.p_f)*(a_1.p_i)*(k.p_i)^2*(k.s_f)*\lambda_f*\lambda_i*|p_i|*c^4
      *\omega-\cos(2 \phi_0)*(a_1.p_f)*(a_1.s_i)*(k.p_f)^2*(k.p_i)*(k.s_f)*\lambda_f*\lambda_i*c^6+
      2.0*\cos(2 \phi_0)*(a_1.p_f)*(a_1.s_i)*(k.p_f)^2*(k.p_i)*\lambda_f*\lambda_i*|p_f|*c^4*\omega
      +2.0*\cos(2 \phi_0)*(a_1.p_f)*(a_1.s_i)*(k.p_f)*(k.p_i)^2*(k.s_f)*\lambda_f*\lambda_i*c^6-
      2.0*\cos(2 \phi_0)*(a_1.p_f)*(a_1.s_i)*(k.p_f)*(k.p_i)^2*\lambda_f*\lambda_i*|p_f|*c^4*\omega
      -2.0*\cos(2 \phi_0)*(a_1.p_f)*(a_1.s_i)*(k.p_f)*(k.p_i)*(k.s_f)*\lambda_f*\lambda_i*c^4*E_i*
      \omega+2.0*\cos(2 \phi_0)*(a_1.p_f)*(a_1.s_i)*(k.p_f)*(k.p_i)*\lambda_f*\lambda_i*|p_f|*c^2*
      E_i*\omega^2-2.0*\cos(2 \phi_0)*(a_1.p_f)*(a_1.s_i)*(k.p_f)*(k.p_i)*\lambda_f*\lambda_i*|p_i|
      *c^2*\cos(\theta_{if})*E_f*\omega^2-\cos(2 \phi_0)*(a_1.p_f)*(a_1.s_i)*(k.p_i)^3*(k.s_f)*
      \lambda_f*\lambda_i*c^6+2.0*\cos(2 \phi_0)*(a_1.p_f)*(a_1.s_i)*(k.p_i)^2*(k.s_f)*\lambda_f*\lambda_i*c^4*E_i
      *\omega-\cos(2 \phi_0)*(a_1.p_i)*(a_1.s_f)*(k.p_f)^3*(k.s_i)*\lambda_f*\lambda_i*c^6+2.0*
      \cos(2 \phi_0)*(a_1.p_i)*(a_1.s_f)*(k.p_f)^2*(k.p_i)*(k.s_i)*\lambda_f*\lambda_i*c^6-2.0*\cos
      (2\phi_0)*(a_1.p_i)*(a_1.s_f)*(k.p_f)^2*(k.p_i)*\lambda_f*\lambda_i*|p_i|*c^4*\omega+2.0*
      \cos(2 \phi_0)*(a_1.p_i)*(a_1.s_f)*(k.p_f)^2*(k.s_i)*\lambda_f*\lambda_i*c^4*E_f*\omega-\cos
      (2\phi_0)*(a_1.p_i)*(a_1.s_f)*(k.p_f)*(k.p_i)^2*(k.s_i)*\lambda_f*\lambda_i*c^6+2.0*\cos(
      2\phi_0)*(a_1.p_i)*(a_1.s_f)*(k.p_f)*(k.p_i)^2*\lambda_f*\lambda_i*|p_i|*c^4*\omega-2.0*\cos
      (2\phi_0)*(a_1.p_i)*(a_1.s_f)*(k.p_f)*(k.p_i)*(k.s_i)*\lambda_f*\lambda_i*c^4*E_f*\omega-2.0*
      \cos(2 \phi_0)*(a_1.p_i)*(a_1.s_f)*(k.p_f)*(k.p_i)*\lambda_f*\lambda_i*|p_f|*c^2*\cos(\theta_{if})*E_i*
      \omega^2+2.0*\cos(2 \phi_0)*(a_1.p_i)*(a_1.s_f)*(k.p_f)*(k.p_i)*\lambda_f*\lambda_i*|p_i|*c^2
      *E_f*\omega^2+\cos(2 \phi_0)*(a_1.s_f)*(a_1.s_i)*(k.p_f)^3*(k.p_i)*\lambda_f*\lambda_i*c^6
      -2.0*\cos(2 \phi_0)*(a_1.s_f)*(a_1.s_i)*(k.p_f)^2*(k.p_i)^2*\lambda_f*\lambda_i*c^6-2.0*
      \cos(2 \phi_0)*(a_1.s_f)*(a_1.s_i)*(k.p_f)^2*(k.p_i)*\lambda_f*\lambda_i*c^4*E_f*\omega+2.0
      *\cos(2 \phi_0)*(a_1.s_f)*(a_1.s_i)*(k.p_f)^2*(k.p_i)*\lambda_f*\lambda_i*c^4*E_i*\omega+
      \cos(2 \phi_0)*(a_1.s_f)*(a_1.s_i)*(k.p_f)*(k.p_i)^3*\lambda_f*\lambda_i*c^6+2.0*\cos(2.0\phi_0)*(a_1.s_f)*(a_1.s_i)*(k.p_f)*(k.p_i)^2*\lambda_f*\lambda_i*c^4*E_f*\omega-2.0*cos(2.0
      \phi_0)*(a_1.s_f)*(a_1.s_i)*(k.p_f)*(k.p_i)^2*\lambda_f*\lambda_i*c^4*E_i*\omega+2.0*\cos(
      2.0*\phi_0 )*(a_1.s_f)*(a_1.s_i)*(k.p_f)*(k.p_i)*\lambda_f*\lambda_i*|p_f|*|p_i|*c^4*\cos(\theta_{if})*\omega
      ^2+2.0*\cos(2 \phi_0)*(a_1.s_f)*(a_1.s_i)*(k.p_f)*(k.p_i)*\lambda_f*\lambda_i*c^6*\omega^2
      -2.0*\cos(2 \phi_0)*(a_1.s_f)*(a_1.s_i)*(k.p_f)*(k.p_i)*\lambda_f*\lambda_i*c^2*E_f*E_i*
      \omega^2-\cos(2 \phi_0)*(a_2.p_f)*(a_2.p_i)*(k.p_f)^2*(k.s_f)*(k.s_i)*\lambda_f*\lambda_i*c^6
      +2.0*\cos(2 \phi_0)*(a_2.p_f)*(a_2.p_i)*(k.p_f)^2*(k.s_i)*\lambda_f*\lambda_i*|p_f|*c^4*
      \omega+2.0*\cos(2 \phi_0)*(a_2.p_f)*(a_2.p_i)*(k.p_f)*(k.p_i)*(k.s_f)*(k.s_i)*\lambda_f*\lambda_i*c
      ^6-2.0*\cos(2 \phi_0)*(a_2.p_f)*(a_2.p_i)*(k.p_f)*(k.p_i)*(k.s_f)*\lambda_f*\lambda_i*|p_i|*c^
      4*\omega-2.0*\cos(2 \phi_0)*(a_2.p_f)*(a_2.p_i)*(k.p_f)*(k.p_i)*(k.s_i)*\lambda_f*\lambda_i*|p_f|*
      c^4*\omega+2.0*\cos(2 \phi_0)*(a_2.p_f)*(a_2.p_i)*(k.p_f)*(k.p_i)*\lambda_f*\lambda_i*|p_f|*
      |p_i|*c^2*\omega^2-2.0*\cos(2 \phi_0)*(a_2.p_f)*(a_2.p_i)*(k.p_f)*(k.p_i)*\lambda_f*\lambda_i
      *\cos(\theta_{if})*E_f*E_i*\omega^2-2.0*\cos(2 \phi_0)*(a_2.p_f)*(a_2.p_i)*(k.p_f)*(k.p_i)*
      c^4*\omega^2-\cos(2 \phi_0)*(a_2.p_f)*(a_2.p_i)*(k.p_i)^2*(k.s_f)*(k.s_i)*\lambda_f*\lambda_i
      *c^6+2.0*\cos(2 \phi_0)*(a_2.p_f)*(a_2.p_i)*(k.p_i)^2*(k.s_f)*\lambda_f*\lambda_i*|p_i|*c^
      4*\omega+\cos(2 \phi_0)*(a_2.p_f)*(a_2.s_i)*(k.p_f)^2*(k.p_i)*(k.s_f)*\lambda_f*\lambda_i*c^6-
      2.0*\cos(2 \phi_0)*(a_2.p_f)*(a_2.s_i)*(k.p_f)^2*(k.p_i)*\lambda_f*\lambda_i*|p_f|*c^4*\omega
      -2.0*\cos(2 \phi_0)*(a_2.p_f)*(a_2.s_i)*(k.p_f)*(k.p_i)^2*(k.s_f)*\lambda_f*\lambda_i*c^6+
      2.0*\cos(2 \phi_0)*(a_2.p_f)*(a_2.s_i)*(k.p_f)*(k.p_i)^2*\lambda_f*\lambda_i*|p_f|*c^4*\omega
      +2.0*\cos(2 \phi_0)*(a_2.p_f)*(a_2.s_i)*(k.p_f)*(k.p_i)*(k.s_f)*\lambda_f*\lambda_i*c^4*E_i*
      \omega-2.0*\cos(2 \phi_0)*(a_2.p_f)*(a_2.s_i)*(k.p_f)*(k.p_i)*\lambda_f*\lambda_i*|p_f|*c^2*
      E_i*\omega^2+2.0*\cos(2 \phi_0)*(a_2.p_f)*(a_2.s_i)*(k.p_f)*(k.p_i)*\lambda_f*\lambda_i*|p_i|*c^2*
      \cos(\theta_{if})*E_f*\omega^2+\cos(2 \phi_0)*(a_2.p_f)*(a_2.s_i)*(k.p_i)^3*(k.s_f)*\lambda_f*\lambda_i
      *c^6-2.0*\cos(2 \phi_0)*(a_2.p_f)*(a_2.s_i)*(k.p_i)^2*(k.s_f)*\lambda_f*\lambda_i*c^4*
      E_i*\omega+\cos(2 \phi_0)*(a_2.p_i)*(a_2.s_f)*(k.p_f)^3*(k.s_i)*\lambda_f*\lambda_i*c^6-
      2.0*\cos(2 \phi_0)*(a_2.p_i)*(a_2.s_f)*(k.p_f)^2*(k.p_i)*(k.s_i)*\lambda_f*\lambda_i*c^6+2.0
      *\cos(2 \phi_0)*(a_2.p_i)*(a_2.s_f)*(k.p_f)^2*(k.p_i)*\lambda_f*\lambda_i*|p_i|*c^4*\omega-
      2.0*\cos(2 \phi_0)*(a_2.p_i)*(a_2.s_f)*(k.p_f)^2*(k.s_i)*\lambda_f*\lambda_i*c^4*E_f*\omega
      +\cos(2 \phi_0)*(a_2.p_i)*(a_2.s_f)*(k.p_f)*(k.p_i)^2*(k.s_i)*\lambda_f*\lambda_i*c^6-2.0*
      \cos(2 \phi_0)*(a_2.p_i)*(a_2.s_f)*(k.p_f)*(k.p_i)^2*\lambda_f*\lambda_i*|p_i|*c^4*\omega+2.0
      *\cos(2 \phi_0)*(a_2.p_i)*(a_2.s_f)*(k.p_f)*(k.p_i)*(k.s_i)*\lambda_f*\lambda_i*c^4*E_f*\omega+
      2.0*\cos(2 \phi_0)*(a_2.p_i)*(a_2.s_f)*(k.p_f)*(k.p_i)*\lambda_f*\lambda_i*|p_f|*c^2*\cos(\theta_{if})*
      E_i*\omega^2-2.0*\cos(2 \phi_0)*(a_2.p_i)*(a_2.s_f)*(k.p_f)*(k.p_i)*\lambda_f*\lambda_i*|p_i|
      *c^2*E_f*\omega^2-\cos(2 \phi_0)*(a_2.s_f)*(a_2.s_i)*(k.p_f)^3*(k.p_i)*\lambda_f*\lambda_i
      *c^6+2.0*\cos(2 \phi_0)*(a_2.s_f)*(a_2.s_i)*(k.p_f)^2*(k.p_i)^2*\lambda_f*\lambda_i*c^6
      +2.0*\cos(2 \phi_0)*(a_2.s_f)*(a_2.s_i)*(k.p_f)^2*(k.p_i)*\lambda_f*\lambda_i*c^4*E_f*
      \omega-2.0*\cos(2 \phi_0)*(a_2.s_f)*(a_2.s_i)*(k.p_f)^2*(k.p_i)*\lambda_f*\lambda_i*c^4*
      E_i*\omega-\cos(2 \phi_0)*(a_2.s_f)*(a_2.s_i)*(k.p_f)*(k.p_i)^3*\lambda_f*\lambda_i*c^6-
      2.0*\cos(2 \phi_0)*(a_2.s_f)*(a_2.s_i)*(k.p_f)*(k.p_i)^2*\lambda_f*\lambda_i*c^4*E_f*\omega
      +2.0*\cos(2 \phi_0)*(a_2.s_f)*(a_2.s_i)*(k.p_f)*(k.p_i)^2*\lambda_f*\lambda_i*c^4*E_i
      *\omega-2.0*\cos(2 \phi_0)*(a_2.s_f)*(a_2.s_i)*(k.p_f)*(k.p_i)*\lambda_f*\lambda_i*|p_f|*|p_i|*c
      ^4*\cos(\theta_{if})*\omega^2-2.0*\cos(2 \phi_0)*(a_2.s_f)*(a_2.s_i)*(k.p_f)*(k.p_i)*\lambda_f*
      \lambda_i*c^6*\omega^2+2.0*\cos(2 \phi_0)*(a_2.s_f)*(a_2.s_i)*(k.p_f)*(k.p_i)*\lambda_f*\lambda_i
      *c^2*E_f*E_i*\omega^2+\sin(2\phi_0)*(a_1.p_f)*(a_2.p_i)*(k.p_f)^2*(k.s_f)*
      (k.s_i)*\lambda_f*\lambda_i*c^6-2.0*\sin(2\phi_0)*(a_1.p_f)*(a_2.p_i)*(k.p_f)^2*(k.s_i)*\lambda_f
      *\lambda_i*|p_f|*c^4*\omega-2.0*\sin(2\phi_0)*(a_1.p_f)*(a_2.p_i)*(k.p_f)*(k.p_i)*(k.s_f)*
      (k.s_i)*\lambda_f*\lambda_i*c^6+2.0*\sin(2\phi_0)*(a_1.p_f)*(a_2.p_i)*(k.p_f)*(k.p_i)*(k.s_f)*
      \lambda_f*\lambda_i*|p_i|*c^4*\omega+2.0*\sin(2\phi_0)*(a_1.p_f)*(a_2.p_i)*(k.p_f)*(k.p_i)*
      (k.s_i)*\lambda_f*\lambda_i*|p_f|*c^4*\omega-2.0*\sin(2\phi_0)*(a_1.p_f)*(a_2.p_i)*(k.p_f)*
      (k.p_i)*\lambda_f*\lambda_i*|p_f|*|p_i|*c^2*\omega^2+2.0*\sin(2\phi_0)*(a_1.p_f)*(a_2.p_i)*
      (k.p_f)*(k.p_i)*\lambda_f*\lambda_i*\cos(\theta_{if})*E_f*E_i*\omega^2+2.0*\sin(2\phi_0)*
      (a_1.p_f)*(a_2.p_i)*(k.p_f)*(k.p_i)*c^4*\omega^2+\sin(2\phi_0)*(a_1.p_f)*(a_2.p_i)*(k.p_i)
      ^2*(k.s_f)*(k.s_i)*\lambda_f*\lambda_i*c^6-2.0*\sin(2\phi_0)*(a_1.p_f)*(a_2.p_i)*(k.p_i)^2
      *(k.s_f)*\lambda_f*\lambda_i*|p_i|*c^4*\omega-\sin(2\phi_0)*(a_1.p_f)*(a_2.s_i)*(k.p_f)^2*
      (k.p_i)*(k.s_f)*\lambda_f*\lambda_i*c^6+2.0*\sin(2\phi_0)*(a_1.p_f)*(a_2.s_i)*(k.p_f)^2*
      (k.p_i)*\lambda_f*\lambda_i*|p_f|*c^4*\omega+2.0*\sin(2\phi_0)*(a_1.p_f)*(a_2.s_i)*(k.p_f)*
      (k.p_i)^2*(k.s_f)*\lambda_f*\lambda_i*c^6-2.0*\sin(2\phi_0)*(a_1.p_f)*(a_2.s_i)*(k.p_f)*
      (k.p_i)^2*\lambda_f*\lambda_i*|p_f|*c^4*\omega-2.0*\sin(2\phi_0)*(a_1.p_f)*(a_2.s_i)*(k.p_f)*(k.p_i)*(k.s_f)*\lambda_f*\lambda_i*c^4*
      E_i*\omega+2.0*\sin(2\phi_0)*(a_1.p_f)*(a_2.s_i)*(k.p_f)*(k.p_i)*\lambda_f*\lambda_i*|p_f|*c
      ^2*E_i*\omega^2-2.0*\sin(2\phi_0)*(a_1.p_f)*(a_2.s_i)*(k.p_f)*(k.p_i)*\lambda_f*\lambda_i
      *|p_i|*c^2*\cos(\theta_{if})*E_f*\omega^2-\sin(2\phi_0)*(a_1.p_f)*(a_2.s_i)*(k.p_i)^3*
      (k.s_f)*\lambda_f*\lambda_i*c^6+2.0*\sin(2\phi_0)*(a_1.p_f)*(a_2.s_i)*(k.p_i)^2*(k.s_f)*\lambda_f
      *\lambda_i*c^4*E_i*\omega+\sin(2\phi_0)*(a_1.p_i)*(a_2.p_f)*(k.p_f)^2*(k.s_f)*(k.s_i)*
      \lambda_f*\lambda_i*c^6-2.0*\sin(2\phi_0)*(a_1.p_i)*(a_2.p_f)*(k.p_f)^2*(k.s_i)*\lambda_f*\lambda_i*
      |p_f|*c^4*\omega-2.0*\sin(2\phi_0)*(a_1.p_i)*(a_2.p_f)*(k.p_f)*(k.p_i)*(k.s_f)*(k.s_i)*
      \lambda_f*\lambda_i*c^6+2.0*\sin(2\phi_0)*(a_1.p_i)*(a_2.p_f)*(k.p_f)*(k.p_i)*(k.s_f)*\lambda_f*
      \lambda_i*|p_i|*c^4*\omega+2.0*\sin(2\phi_0)*(a_1.p_i)*(a_2.p_f)*(k.p_f)*(k.p_i)*(k.s_i)*
      \lambda_f*\lambda_i*|p_f|*c^4*\omega-2.0*\sin(2\phi_0)*(a_1.p_i)*(a_2.p_f)*(k.p_f)*(k.p_i)*
      \lambda_f*\lambda_i*|p_f|*|p_i|*c^2*\omega^2+2.0*\sin(2\phi_0)*(a_1.p_i)*(a_2.p_f)*(k.p_f)*
      (k.p_i)*\lambda_f*\lambda_i*\cos(\theta_{if})*E_f*E_i*\omega^2+2.0*\sin(2\phi_0)*(a_1.p_i)*
      (a_2.p_f)*(k.p_f)*(k.p_i)*c^4*\omega^2+\sin(2\phi_0)*(a_1.p_i)*(a_2.p_f)*(k.p_i)^2*
      (k.s_f)*(k.s_i)*\lambda_f*\lambda_i*c^6-2.0*\sin(2\phi_0)*(a_1.p_i)*(a_2.p_f)*(k.p_i)^2*
      (k.s_f)*\lambda_f*\lambda_i*|p_i|*c^4*\omega-\sin(2\phi_0)*(a_1.p_i)*(a_2.s_f)*(k.p_f)^3*
      (k.s_i)*\lambda_f*\lambda_i*c^6+2.0*\sin(2\phi_0)*(a_1.p_i)*(a_2.s_f)*(k.p_f)^2*(k.p_i)*
      (k.s_i)*\lambda_f*\lambda_i*c^6-2.0*\sin(2\phi_0)*(a_1.p_i)*(a_2.s_f)*(k.p_f)^2*(k.p_i)*\lambda_f
      *\lambda_i*|p_i|*c^4*\omega+2.0*\sin(2\phi_0)*(a_1.p_i)*(a_2.s_f)*(k.p_f)^2*(k.s_i)*\lambda_f*\lambda_i*c^4*E_f*
      \omega-\sin(2\phi_0)*(a_1.p_i)*(a_2.s_f)*(k.p_f)*(k.p_i)^2*(k.s_i)*\lambda_f*\lambda_i*c^6+
      2.0*\sin(2\phi_0)*(a_1.p_i)*(a_2.s_f)*(k.p_f)*(k.p_i)^2*\lambda_f*\lambda_i*|p_i|*c^4*\omega
      -2.0*\sin(2\phi_0)*(a_1.p_i)*(a_2.s_f)*(k.p_f)*(k.p_i)*(k.s_i)*\lambda_f*\lambda_i*c^4*E_f*
      \omega-2.0*\sin(2\phi_0)*(a_1.p_i)*(a_2.s_f)*(k.p_f)*(k.p_i)*\lambda_f*\lambda_i*|p_f|*c^2*
      \cos(\theta_{if})*E_i*\omega^2+2.0*\sin(2\phi_0)*(a_1.p_i)*(a_2.s_f)*(k.p_f)*(k.p_i)*\lambda_f*
      \lambda_i*|p_i|*c^2*E_f*\omega^2-\sin(2\phi_0)*(a_1.s_f)*(a_2.p_i)*(k.p_f)^3*(k.s_i)*
      \lambda_f*\lambda_i*c^6+2.0*\sin(2\phi_0)*(a_1.s_f)*(a_2.p_i)*(k.p_f)^2*(k.p_i)*(k.s_i)*\lambda_f
      *\lambda_i*c^6-2.0*\sin(2\phi_0)*(a_1.s_f)*(a_2.p_i)*(k.p_f)^2*(k.p_i)*\lambda_f*\lambda_i*|p_i|
      *c^4*\omega+2.0*\sin(2\phi_0)*(a_1.s_f)*(a_2.p_i)*(k.p_f)^2*(k.s_i)*\lambda_f*\lambda_i*c
      ^4*E_f*\omega-\sin(2\phi_0)*(a_1.s_f)*(a_2.p_i)*(k.p_f)*(k.p_i)^2*(k.s_i)*\lambda_f*\lambda_i
      *c^6+2.0*\sin(2\phi_0)*(a_1.s_f)*(a_2.p_i)*(k.p_f)*(k.p_i)^2*\lambda_f*\lambda_i*|p_i|*c^
      4*\omega-2.0*\sin(2\phi_0)*(a_1.s_f)*(a_2.p_i)*(k.p_f)*(k.p_i)*(k.s_i)*\lambda_f*\lambda_i*c^4
      *E_f*\omega-2.0*\sin(2\phi_0)*(a_1.s_f)*(a_2.p_i)*(k.p_f)*(k.p_i)*\lambda_f*\lambda_i*|p_f|*c
      ^2*\cos(\theta_{if})*E_i*\omega^2+2.0*\sin(2\phi_0)*(a_1.s_f)*(a_2.p_i)*(k.p_f)*(k.p_i)*
      \lambda_f*\lambda_i*|p_i|*c^2*E_f*\omega^2+\sin(2\phi_0)*(a_1.s_f)*(a_2.s_i)*(k.p_f)^3*
      (k.p_i)*\lambda_f*\lambda_i*c^6-2.0*\sin(2\phi_0)*(a_1.s_f)*(a_2.s_i)*(k.p_f)^2*(k.p_i)^2*
      \lambda_f*\lambda_i*c^6-2.0*\sin(2\phi_0)*(a_1.s_f)*(a_2.s_i)*(k.p_f)^2*(k.p_i)*\lambda_f*\lambda_i*
      c^4*E_f*\omega+2.0*\sin(2\phi_0)*(a_1.s_f)*(a_2.s_i)*(k.p_f)^2*(k.p_i)*\lambda_f*\lambda_i*c^4*E_i
      *\omega+\sin(2\phi_0)*(a_1.s_f)*(a_2.s_i)*(k.p_f)*(k.p_i)^3*\lambda_f*\lambda_i*c^6+2.0*
      \sin(2\phi_0)*(a_1.s_f)*(a_2.s_i)*(k.p_f)*(k.p_i)^2*\lambda_f*\lambda_i*c^4*E_f*\omega-2.0
      *\sin(2\phi_0)*(a_1.s_f)*(a_2.s_i)*(k.p_f)*(k.p_i)^2*\lambda_f*\lambda_i*c^4*E_i*\omega+
      2.0*\sin(2\phi_0)*(a_1.s_f)*(a_2.s_i)*(k.p_f)*(k.p_i)*\lambda_f*\lambda_i*|p_f|*|p_i|*c^4*
      \cos(\theta_{if})*\omega^2+2.0*\sin(2\phi_0)*(a_1.s_f)*(a_2.s_i)*(k.p_f)*(k.p_i)*\lambda_f*\lambda_i*c
      ^6*\omega^2-2.0*\sin(2\phi_0)*(a_1.s_f)*(a_2.s_i)*(k.p_f)*(k.p_i)*\lambda_f*\lambda_i*c^2
      *E_f*E_i*\omega^2-\sin(2\phi_0)*(a_1.s_i)*(a_2.p_f)*(k.p_f)^2*(k.p_i)*(k.s_f)*
      \lambda_f*\lambda_i*c^6+2.0*\sin(2\phi_0)*(a_1.s_i)*(a_2.p_f)*(k.p_f)^2*(k.p_i)*\lambda_f*\lambda_i*
      |p_f|*c^4*\omega+2.0*\sin(2\phi_0)*(a_1.s_i)*(a_2.p_f)*(k.p_f)*(k.p_i)^2*(k.s_f)*
      \lambda_f*\lambda_i*c^6-2.0*\sin(2\phi_0)*(a_1.s_i)*(a_2.p_f)*(k.p_f)*(k.p_i)^2*\lambda_f*\lambda_i*
      |p_f|*c^4*\omega-2.0*\sin(2\phi_0)*(a_1.s_i)*(a_2.p_f)*(k.p_f)*(k.p_i)*(k.s_f)*\lambda_f*
      \lambda_i*c^4*E_i*\omega+2.0*\sin(2\phi_0)*(a_1.s_i)*(a_2.p_f)*(k.p_f)*(k.p_i)*\lambda_f*
      \lambda_i*|p_f|*c^2*E_i*\omega^2-2.0*\sin(2\phi_0)*(a_1.s_i)*(a_2.p_f)*(k.p_f)*
      (k.p_i)*\lambda_f*\lambda_i*|p_i|*c^2*\cos(\theta_{if})*E_f*\omega^2-\sin(2\phi_0)*(a_1.s_i)*
      (a_2.p_f)*(k.p_i)^3*(k.s_f)*\lambda_f*\lambda_i*c^6+2.0*\sin(2\phi_0)*(a_1.s_i)*(a_2.p_f)*
      (k.p_i)^2*(k.s_f)*\lambda_f*\lambda_i*c^4*E_i*\omega+\sin(2\phi_0)*(a_1.s_i)*(a_2.s_f)*
      (k.p_f)^3*(k.p_i)*\lambda_f*\lambda_i*c^6-2.0*\sin(2\phi_0)*(a_1.s_i)*(a_2.s_f)*(k.p_f)^2*
      (k.p_i)^2*\lambda_f*\lambda_i*c^6-2.0*\sin(2\phi_0)*(a_1.s_i)*(a_2.s_f)*(k.p_f)^2*(k.p_i)*\lambda_f*\lambda_i*c^4*E_f
      *\omega+2.0*\sin(2\phi_0)*(a_1.s_i)*(a_2.s_f)*(k.p_f)^2*(k.p_i)*\lambda_f*\lambda_i*c^4*
      E_i*\omega+\sin(2\phi_0)*(a_1.s_i)*(a_2.s_f)*(k.p_f)*(k.p_i)^3*\lambda_f*\lambda_i*c^6+
      2.0*\sin(2\phi_0)*(a_1.s_i)*(a_2.s_f)*(k.p_f)*(k.p_i)^2*\lambda_f*\lambda_i*c^4*E_f*\omega
      -2.0*\sin(2\phi_0)*(a_1.s_i)*(a_2.s_f)*(k.p_f)*(k.p_i)^2*\lambda_f*\lambda_i*c^4*E_i*
      \omega+2.0*\sin(2\phi_0)*(a_1.s_i)*(a_2.s_f)*(k.p_f)*(k.p_i)*\lambda_f*\lambda_i*|p_f|*|p_i|*c^
      4*\cos(\theta_{if})*\omega^2+2.0*\sin(2\phi_0)*(a_1.s_i)*(a_2.s_f)*(k.p_f)*(k.p_i)*\lambda_f*\lambda_i*
      c^6*\omega^2-2.0*\sin(2\phi_0)*(a_1.s_i)*(a_2.s_f)*(k.p_f)*(k.p_i)*\lambda_f*\lambda_i*c^
      2*E_f*E_i*\omega^2\Big\}/[2.0*(k.p_f)^2*(k.p_i)^2*c^6]
$

\end{sloppypar}

\begin{sloppypar}
$\mathcal{D}(\lambda_i,\lambda_f)=\Big\{\cos(\phi_0)*(a_1.p_f)*(k.p_f)^2*(k.p_i)^2*\lambda_f*\lambda_i*|p_f|*|p_i|*c^6+cos(
      \phi_0 )*(a_1.p_f)*(k.p_f)^2*(k.p_i)^2*\lambda_f*\lambda_i*c^4*\cos(\theta_{if})*E_f*E_i+cos(
      \phi_0 )*(a_1.p_f)*(k.p_f)^2*(k.p_i)^2*c^8-\cos(\phi_0)*(a_1.p_f)*(k.p_f)^2*(k.p_i)*
      (k.s_i)*\lambda_f*\lambda_i*|p_f|*c^6*E_i-\cos(\phi_0)*(a_1.p_f)*(k.p_f)^2*(k.p_i)*(k.s_i)*
      \lambda_f*\lambda_i*|p_i|*c^6*\cos(\theta_{if})*E_f-\cos(\phi_0)*(a_1.p_f)*(k.p_f)*(k.p_i)^3*\lambda_f*\lambda_i
      *|p_f|*|p_i|*c^6-\cos(\phi_0)*(a_1.p_f)*(k.p_f)*(k.p_i)^3*\lambda_f*\lambda_i*c^4*\cos(\theta_{if})*
      E_f*E_i-\cos(\phi_0)*(a_1.p_f)*(k.p_f)*(k.p_i)^3*c^8+\cos(\phi_0)*(a_1.p_f)*(k.p_f)*
      (k.p_i)^2*(k.s_i)*\lambda_f*\lambda_i*|p_f|*c^6*E_i+\cos(\phi_0)*(a_1.p_f)*(k.p_f)*(k.p_i)^2
      *(k.s_i)*\lambda_f*\lambda_i*|p_i|*c^6*\cos(\theta_{if})*E_f-2.0*\cos(\phi_0)*(a_1.p_f)*(k.p_f)*(k.p_i)
      ^2*\lambda_f*\lambda_i*|p_i|^2*c^4*\cos(\theta_{if})*E_f*\omega+2.0*\cos(\phi_0)*(a_1.p_f)*(k.p_f)
      *(k.p_i)^2*\lambda_f*\lambda_i*c^2*\cos(\theta_{if})*E_f*E_i^2*\omega+2.0*\cos(\phi_0)*
      (a_1.p_f)*(k.p_f)*(k.p_i)^2*c^6*E_i*\omega-\cos(\phi_0)*(a_1.p_f)*(k.p_f)*(k.p_i)*(k.s_f)
      *\lambda_f*\lambda_i*(a^2)*|p_i|*c^4*\omega+\cos(\phi_0)*(a_1.p_f)*(k.p_f)*(k.p_i)*\lambda_f*\lambda_i*
      (a^2)*|p_f|*|p_i|*c^2*\omega^2-\cos(\phi_0)*(a_1.p_f)*(k.p_f)*(k.p_i)*\lambda_f*\lambda_i*(a^2)*
      \cos(\theta_{if})*E_f*E_i*\omega^2-\cos(\phi_0)*(a_1.p_f)*(k.p_f)*(k.p_i)*(a^2)*c^4*\omega
      ^2+\cos(\phi_0)*(a_1.p_f)*(k.p_f)*(k.s_f)*(k.s_i)*\lambda_f*\lambda_i*(a^2)*c^4*E_i*\omega-
      \cos(\phi_0)*(a_1.p_f)*(k.p_f)*(k.s_i)*\lambda_f*\lambda_i*(a^2)*|p_f|*c^2*E_i*\omega^2+cos
      (\phi_0 )*(a_1.p_f)*(k.p_f)*(k.s_i)*\lambda_f*\lambda_i*(a^2)*|p_i|*c^2*\cos(\theta_{if})*E_f*\omega^2+\cos(\phi_0)*(a_1.p_f)*(k.p_i)^2*(k.s_f)*\lambda_f*\lambda_i*(a^2)*|p_i|*c^4*\omega-cos(
      \phi_0 )*(a_1.p_f)*(k.p_i)*(k.s_f)*(k.s_i)*\lambda_f*\lambda_i*(a^2)*c^4*E_i*\omega-\cos(\phi_0)
      *(a_1.p_i)*(k.p_f)^3*(k.p_i)*\lambda_f*\lambda_i*|p_f|*|p_i|*c^6-\cos(\phi_0)*(a_1.p_i)*(k.p_f)^3
      *(k.p_i)*\lambda_f*\lambda_i*c^4*\cos(\theta_{if})*E_f*E_i-\cos(\phi_0)*(a_1.p_i)*(k.p_f)^3*(k.p_i)*
      c^8+\cos(\phi_0)*(a_1.p_i)*(k.p_f)^2*(k.p_i)^2*\lambda_f*\lambda_i*|p_f|*|p_i|*c^6+cos(
      \phi_0 )*(a_1.p_i)*(k.p_f)^2*(k.p_i)^2*\lambda_f*\lambda_i*c^4*\cos(\theta_{if})*E_f*E_i+cos(
      \phi_0 )*(a_1.p_i)*(k.p_f)^2*(k.p_i)^2*c^8+\cos(\phi_0)*(a_1.p_i)*(k.p_f)^2*(k.p_i)*
      (k.s_f)*\lambda_f*\lambda_i*|p_f|*c^6*\cos(\theta_{if})*E_i+\cos(\phi_0)*(a_1.p_i)*(k.p_f)^2*(k.p_i)*
      (k.s_f)*\lambda_f*\lambda_i*|p_i|*c^6*E_f-2.0*\cos(\phi_0)*(a_1.p_i)*(k.p_f)^2*(k.p_i)*\lambda_f*
      \lambda_i*|p_f|^2*c^4*\cos(\theta_{if})*E_i*\omega+2.0*\cos(\phi_0)*(a_1.p_i)*(k.p_f)^2*
      (k.p_i)*\lambda_f*\lambda_i*c^2*\cos(\theta_{if})*E_f^2*E_i*\omega+2.0*\cos(\phi_0)*(a_1.p_i)*
      (k.p_f)^2*(k.p_i)*c^6*E_f*\omega+\cos(\phi_0)*(a_1.p_i)*(k.p_f)^2*(k.s_i)*\lambda_f*\lambda_i
      *(a^2)*|p_f|*c^4*\omega-\cos(\phi_0)*(a_1.p_i)*(k.p_f)*(k.p_i)^2*(k.s_f)*\lambda_f*\lambda_i*
      |p_f|*c^6*\cos(\theta_{if})*E_i-\cos(\phi_0)*(a_1.p_i)*(k.p_f)*(k.p_i)^2*(k.s_f)*\lambda_f*\lambda_i*
      |p_i|*c^6*E_f-\cos(\phi_0)*(a_1.p_i)*(k.p_f)*(k.p_i)*(k.s_i)*\lambda_f*\lambda_i*(a^2)*|p_f|*c^
      4*\omega+\cos(\phi_0)*(a_1.p_i)*(k.p_f)*(k.p_i)*\lambda_f*\lambda_i*(a^2)*|p_f|*|p_i|*c^2*\omega
      ^2-\cos(\phi_0)*(a_1.p_i)*(k.p_f)*(k.p_i)*\lambda_f*\lambda_i*(a^2)*\cos(\theta_{if})*E_f*E_i*\omega^
      2-\cos(\phi_0)*(a_1.p_i)*(k.p_f)*(k.p_i)*(a^2)*c^4*\omega^2-\cos(\phi_0)*(a_1.p_i)*(k.p_f)*(k.s_f)*(k.s_i)*\lambda_f*\lambda_i*(a^2)*c^4*E_f*\omega+
      \cos(\phi_0)*(a_1.p_i)*(k.p_i)*(k.s_f)*(k.s_i)*\lambda_f*\lambda_i*(a^2)*c^4*E_f*\omega+cos(
      \phi_0 )*(a_1.p_i)*(k.p_i)*(k.s_f)*\lambda_f*\lambda_i*(a^2)*|p_f|*c^2*\cos(\theta_{if})*E_i*\omega^2-
      \cos(\phi_0)*(a_1.p_i)*(k.p_i)*(k.s_f)*\lambda_f*\lambda_i*(a^2)*|p_i|*c^2*E_f*\omega^2-cos(
      \phi_0 )*(a_1.s_f)*(k.p_f)^2*(k.p_i)^2*\lambda_f*\lambda_i*|p_f|*c^6*\cos(\theta_{if})*E_i-cos(
      \phi_0 )*(a_1.s_f)*(k.p_f)^2*(k.p_i)^2*\lambda_f*\lambda_i*|p_i|*c^6*E_f+\cos(\phi_0)*(a_1.s_f)
      *(k.p_f)^2*(k.p_i)*(k.s_i)*\lambda_f*\lambda_i*|p_f|*|p_i|*c^8*\cos(\theta_{if})+\cos(\phi_0)*(a_1.s_f)*
      (k.p_f)^2*(k.p_i)*(k.s_i)*\lambda_f*\lambda_i*c^{10}+\cos(\phi_0)*(a_1.s_f)*(k.p_f)^2*(k.p_i)*
      (k.s_i)*\lambda_f*\lambda_i*c^6*E_f*E_i+\cos(\phi_0)*(a_1.s_f)*(k.p_f)^2*(k.p_i)*\lambda_f*\lambda_i
      *(a^2)*|p_i|*c^4*\omega-\cos(\phi_0)*(a_1.s_f)*(k.p_f)^2*(k.s_i)*\lambda_f*\lambda_i*(a^2)*c^
      4*E_i*\omega+\cos(\phi_0)*(a_1.s_f)*(k.p_f)*(k.p_i)^3*\lambda_f*\lambda_i*|p_f|*c^6*\cos(\theta_{if})
      *E_i+\cos(\phi_0)*(a_1.s_f)*(k.p_f)*(k.p_i)^3*\lambda_f*\lambda_i*|p_i|*c^6*E_f-cos(
      \phi_0 )*(a_1.s_f)*(k.p_f)*(k.p_i)^2*(k.s_i)*\lambda_f*\lambda_i*|p_f|*|p_i|*c^8*\cos(\theta_{if})-cos(
      \phi_0 )*(a_1.s_f)*(k.p_f)*(k.p_i)^2*(k.s_i)*\lambda_f*\lambda_i*c^{10}-\cos(\phi_0)*(a_1.s_f)*(k.p_f)
      *(k.p_i)^2*(k.s_i)*\lambda_f*\lambda_i*c^6*E_f*E_i-\cos(\phi_0)*(a_1.s_f)*(k.p_f)*(k.p_i)^
      2*\lambda_f*\lambda_i*(a^2)*|p_i|*c^4*\omega+2.0*\cos(\phi_0)*(a_1.s_f)*(k.p_f)*(k.p_i)^2*
      \lambda_f*\lambda_i*|p_f|*|p_i|^2*c^6*\cos(\theta_{if})*\omega-2.0*\cos(\phi_0)*(a_1.s_f)*(k.p_f)*
      (k.p_i)^2*\lambda_f*\lambda_i*|p_f|*c^4*\cos(\theta_{if})*E_i^2*\omega+2.0*\cos(\phi_0)*(a_1.s_f)
      *(k.p_f)*(k.p_i)^2*\lambda_f*\lambda_i*|p_i|*c^8*\omega+\cos(\phi_0)*(a_1.s_f)*(k.p_f)*(k.p_i)*(k.s_i)*\lambda_f*\lambda_i*(a^2)*c^4*E_i*\omega+
      \cos(\phi_0)*(a_1.s_f)*(k.p_f)*(k.p_i)*\lambda_f*\lambda_i*(a^2)*|p_f|*c^2*\cos(\theta_{if})*E_i*\omega
      ^2-\cos(\phi_0)*(a_1.s_f)*(k.p_f)*(k.p_i)*\lambda_f*\lambda_i*(a^2)*|p_i|*c^2*E_f*\omega^2-
      \cos(\phi_0)*(a_1.s_f)*(k.p_f)*(k.s_i)*\lambda_f*\lambda_i*(a^2)*|p_f|*|p_i|*c^4*\cos(\theta_{if})*\omega^
      2-\cos(\phi_0)*(a_1.s_f)*(k.p_f)*(k.s_i)*\lambda_f*\lambda_i*(a^2)*c^6*\omega^2+\cos(\phi_0)*
      (a_1.s_f)*(k.p_f)*(k.s_i)*\lambda_f*\lambda_i*(a^2)*c^2*E_f*E_i*\omega^2+\cos(\phi_0)*
      (a_1.s_i)*(k.p_f)^3*(k.p_i)*\lambda_f*\lambda_i*|p_f|*c^6*E_i+\cos(\phi_0)*(a_1.s_i)*(k.p_f)^3
      *(k.p_i)*\lambda_f*\lambda_i*|p_i|*c^6*\cos(\theta_{if})*E_f-\cos(\phi_0)*(a_1.s_i)*(k.p_f)^2*(k.p_i)^
      2*\lambda_f*\lambda_i*|p_f|*c^6*E_i-\cos(\phi_0)*(a_1.s_i)*(k.p_f)^2*(k.p_i)^2*\lambda_f*\lambda_i
      *|p_i|*c^6*\cos(\theta_{if})*E_f-\cos(\phi_0)*(a_1.s_i)*(k.p_f)^2*(k.p_i)*(k.s_f)*\lambda_f*\lambda_i*
      |p_f|*|p_i|*c^8*\cos(\theta_{if})-\cos(\phi_0)*(a_1.s_i)*(k.p_f)^2*(k.p_i)*(k.s_f)*\lambda_f*\lambda_i*c
      ^{10}-\cos(\phi_0)*(a_1.s_i)*(k.p_f)^2*(k.p_i)*(k.s_f)*\lambda_f*\lambda_i*c^6*E_f*E_i-
      \cos(\phi_0)*(a_1.s_i)*(k.p_f)^2*(k.p_i)*\lambda_f*\lambda_i*(a^2)*|p_f|*c^4*\omega+2.0*cos(
      \phi_0 )*(a_1.s_i)*(k.p_f)^2*(k.p_i)*\lambda_f*\lambda_i*|p_f|^2*|p_i|*c^6*\cos(\theta_{if})*\omega+2.0
      *\cos(\phi_0)*(a_1.s_i)*(k.p_f)^2*(k.p_i)*\lambda_f*\lambda_i*|p_f|*c^8*\omega-2.0*cos(
      \phi_0 )*(a_1.s_i)*(k.p_f)^2*(k.p_i)*\lambda_f*\lambda_i*|p_i|*c^4*\cos(\theta_{if})*E_f^2*\omega+cos
      (\phi_0 )*(a_1.s_i)*(k.p_f)*(k.p_i)^2*(k.s_f)*\lambda_f*\lambda_i*|p_f|*|p_i|*c^8*\cos(\theta_{if})+cos(
      \phi_0 )*(a_1.s_i)*(k.p_f)*(k.p_i)^2*(k.s_f)*\lambda_f*\lambda_i*c^{10}+\cos(\phi_0)*(a_1.s_i)*(k.p_f)*(k.p_i)^2*(k.s_f)*\lambda_f*\lambda_i*c^6*E_f*E_i+cos
      (\phi_0 )*(a_1.s_i)*(k.p_f)*(k.p_i)^2*\lambda_f*\lambda_i*(a^2)*|p_f|*c^4*\omega+\cos(\phi_0)*
      (a_1.s_i)*(k.p_f)*(k.p_i)*(k.s_f)*\lambda_f*\lambda_i*(a^2)*c^4*E_f*\omega-\cos(\phi_0)*(a_1.s_i)*
      (k.p_f)*(k.p_i)*\lambda_f*\lambda_i*(a^2)*|p_f|*c^2*E_i*\omega^2+\cos(\phi_0)*(a_1.s_i)*
      (k.p_f)*(k.p_i)*\lambda_f*\lambda_i*(a^2)*|p_i|*c^2*\cos(\theta_{if})*E_f*\omega^2-\cos(\phi_0)*
      (a_1.s_i)*(k.p_i)^2*(k.s_f)*\lambda_f*\lambda_i*(a^2)*c^4*E_f*\omega-\cos(\phi_0)*(a_1.s_i)*
      (k.p_i)*(k.s_f)*\lambda_f*\lambda_i*(a^2)*|p_f|*|p_i|*c^4*\cos(\theta_{if})*\omega^2-\cos(\phi_0)*
      (a_1.s_i)*(k.p_i)*(k.s_f)*\lambda_f*\lambda_i*(a^2)*c^6*\omega^2+\cos(\phi_0)*(a_1.s_i)*(k.p_i)*
      (k.s_f)*\lambda_f*\lambda_i*(a^2)*c^2*E_f*E_i*\omega^2-\cos(\phi_0)*(k.p_f)^2*(k.p_i)*
      \lambda_f*|p_f|^2*c^7*\cos(\varphi_f)*\omega*\sin(\theta_f)*|a|+\cos(\phi_0)*(k.p_f)^2*(k.p_i)*
      \lambda_f*|p_f|*|p_i|*c^7*\cos(\varphi_f)*\omega*\sin(\theta_i)*|a|-\cos(\phi_0)*(k.p_f)^2*(k.p_i)
      *\lambda_f*|p_i|*c^6*\cos(\varphi_f)*\cos(\theta_i)*E_f*\omega*\sin(\theta_f)*|a|+\cos(\phi_0)*(k.p_f)
      ^2*(k.p_i)*\lambda_f*|p_i|*c^6*\cos(\varphi_f)*\cos(\theta_f)*E_f*\omega*\sin(\theta_i)*|a|+cos(
      \phi_0 )*(k.p_f)^2*(k.p_i)*\lambda_f*c^5*\cos(\varphi_f)*E_f^2*\omega*\sin(\theta_f)*|a|+cos(
      \phi_0 )*(k.p_f)^2*(k.p_i)*\lambda_f*c^5*\cos(\varphi_f)*E_f*E_i*\omega*\sin(\theta_f)*|a|+
      \cos(\phi_0)*(k.p_f)^2*(k.p_i)*\lambda_i*|p_f|*|p_i|*c^7*\cos(\varphi_f)*\omega*\sin(\theta_f)*|a|
      -\cos(\phi_0)*(k.p_f)^2*(k.p_i)*\lambda_i*|p_f|*c^6*\cos(\varphi_f)*\cos(\theta_i)*E_i*\omega*
      \sin(\theta_f)*|a|+\cos(\phi_0)*(k.p_f)^2*(k.p_i)*\lambda_i*|p_f|*c^6*\cos(\varphi_f)*\cos(\theta_f)*
      E_i*\omega*\sin(\theta_i)*|a|-\cos(\phi_0)*(k.p_f)^2*(k.p_i)*\lambda_i*|p_i|^2*c^7*\cos(\varphi_f)*\omega*\sin(\theta_i)*
      |a|+\cos(\phi_0)*(k.p_f)^2*(k.p_i)*\lambda_i*c^5*\cos(\varphi_f)*E_f*E_i*\omega*
      \sin(\theta_i)*|a|+\cos(\phi_0)*(k.p_f)^2*(k.p_i)*\lambda_i*c^5*\cos(\varphi_f)*E_i^2*\omega
      *\sin(\theta_i)*|a|+\cos(\phi_0)*(k.p_f)*(k.p_i)^2*\lambda_f*|p_f|^2*c^7*\cos(\varphi_f)*\omega
      *\sin(\theta_f)*|a|-\cos(\phi_0)*(k.p_f)*(k.p_i)^2*\lambda_f*|p_f|*|p_i|*c^7*\cos(\varphi_f)*
      \omega*\sin(\theta_i)*|a|-\cos(\phi_0)*(k.p_f)*(k.p_i)^2*\lambda_f*|p_i|*c^6*\cos(\varphi_f)*
      \cos(\theta_i)*E_f*\omega*\sin(\theta_f)*|a|+\cos(\phi_0)*(k.p_f)*(k.p_i)^2*\lambda_f*|p_i|*c^6
      *\cos(\varphi_f)*\cos(\theta_f)*E_f*\omega*\sin(\theta_i)*|a|-\cos(\phi_0)*(k.p_f)*(k.p_i)^2*\lambda_f*c
      ^5*\cos(\varphi_f)*E_f^2*\omega*\sin(\theta_f)*|a|-\cos(\phi_0)*(k.p_f)*(k.p_i)^2*\lambda_f*c
      ^5*\cos(\varphi_f)*E_f*E_i*\omega*\sin(\theta_f)*|a|-\cos(\phi_0)*(k.p_f)*(k.p_i)^2*\lambda_i
      *|p_f|*|p_i|*c^7*\cos(\varphi_f)*\omega*\sin(\theta_f)*|a|-\cos(\phi_0)*(k.p_f)*(k.p_i)^2*
      \lambda_i*|p_f|*c^6*\cos(\varphi_f)*\cos(\theta_i)*E_i*\omega*\sin(\theta_f)*|a|+\cos(\phi_0)*(k.p_f)*
      (k.p_i)^2*\lambda_i*|p_f|*c^6*\cos(\varphi_f)*\cos(\theta_f)*E_i*\omega*\sin(\theta_i)*|a|+cos(
      \phi_0 )*(k.p_f)*(k.p_i)^2*\lambda_i*|p_i|^2*c^7*\cos(\varphi_f)*\omega*\sin(\theta_i)*|a|-cos(
      \phi_0 )*(k.p_f)*(k.p_i)^2*\lambda_i*c^5*\cos(\varphi_f)*E_f*E_i*\omega*\sin(\theta_i)*|a|-
      \cos(\phi_0)*(k.p_f)*(k.p_i)^2*\lambda_i*c^5*\cos(\varphi_f)*E_i^2*\omega*\sin(\theta_i)*|a|
      -\cos(\phi_0)*(k.p_f)*(k.p_i)*\lambda_f*(a^2)*c^3*\cos(\varphi_f)*E_f*\omega^2*\sin(\theta_f)*
      |a|-\cos(\phi_0)*(k.p_f)*(k.p_i)*\lambda_i*(a^2)*c^3*\cos(\varphi_f)*E_i*\omega^2*
      \sin(\theta_i)*|a|-\cos(\phi_0)*(k.p_f)*(k.s_i)*\lambda_i*(a^2)*|p_f|*c^5*\cos(\varphi_f)*\omega^2*\sin(\theta_f)*
      |a|+\cos(\phi_0)*(k.p_f)*(k.s_i)*\lambda_i*(a^2)*|p_i|*c^5*\cos(\varphi_f)*\omega^2*\sin(\theta_i)
      *|a|+\cos(\phi_0)*(k.p_i)^2*\lambda_f*(a^2)*c^3*\cos(\varphi_f)*E_f*\omega^2*\sin(\theta_f)*
      |a|+\cos(\phi_0)*(k.p_i)^2*\lambda_i*(a^2)*c^3*\cos(\varphi_f)*E_i*\omega^2*\sin(\theta_i)*
      |a|+\cos(\phi_0)*(k.p_i)*(k.s_i)*\lambda_i*(a^2)*|p_f|*c^5*\cos(\varphi_f)*\omega^2*\sin(\theta_f)
      *|a|-\cos(\phi_0)*(k.p_i)*(k.s_i)*\lambda_i*(a^2)*|p_i|*c^5*\cos(\varphi_f)*\omega^2*
      \sin(\theta_i)*|a|-\cos(\phi_0)*(k.p_i)*\lambda_f*(a^2)*|p_f|^2*c^3*\cos(\varphi_f)*\omega^3*
      \sin(\theta_f)*|a|+\cos(\phi_0)*(k.p_i)*\lambda_f*(a^2)*|p_f|*|p_i|*c^3*\cos(\varphi_f)*\omega^3*
      \sin(\theta_i)*|a|+\cos(\phi_0)*(k.p_i)*\lambda_f*(a^2)*|p_i|*c^2*\cos(\varphi_f)*\cos(\theta_i)*E_f*
      \omega^3*\sin(\theta_f)*|a|-\cos(\phi_0)*(k.p_i)*\lambda_f*(a^2)*|p_i|*c^2*\cos(\varphi_f)*
      \cos(\theta_f)*E_f*\omega^3*\sin(\theta_i)*|a|+\cos(\phi_0)*(k.p_i)*\lambda_f*(a^2)*c*\cos(\varphi_f)*
      E_f^2*\omega^3*\sin(\theta_f)*|a|-\cos(\phi_0)*(k.p_i)*\lambda_f*(a^2)*c*\cos(\varphi_f)*E_f*
      E_i*\omega^3*\sin(\theta_f)*|a|-\cos(\phi_0)*(k.p_i)*\lambda_i*(a^2)*|p_f|*|p_i|*c^3*
      \cos(\varphi_f)*\omega^3*\sin(\theta_f)*|a|+\cos(\phi_0)*(k.p_i)*\lambda_i*(a^2)*|p_f|*c^2*
      \cos(\varphi_f)*\cos(\theta_i)*E_i*\omega^3*\sin(\theta_f)*|a|-\cos(\phi_0)*(k.p_i)*\lambda_i*(a^2)*
      |p_f|*c^2*\cos(\varphi_f)*\cos(\theta_f)*E_i*\omega^3*\sin(\theta_i)*|a|+\cos(\phi_0)*(k.p_i)*
      \lambda_i*(a^2)*|p_i|^2*c^3*\cos(\varphi_f)*\omega^3*\sin(\theta_i)*|a|+\cos(\phi_0)*(k.p_i)*
      \lambda_i*(a^2)*c*\cos(\varphi_f)*E_f*E_i*\omega^3*\sin(\theta_i)*|a|-\cos(\phi_0)*(k.p_i)*
      \lambda_i*(a^2)*c*\cos(\varphi_f)*E_i^2*\omega^3*\sin(\theta_i)*|a|+\sin(\phi_0)*(a_2.p_f)*(k.p_f)^2*(k.p_i)^2*\lambda_f*\lambda_i*|p_f|*|p_i|*c^6+\sin(
      \phi_0 )*(a_2.p_f)*(k.p_f)^2*(k.p_i)^2*\lambda_f*\lambda_i*c^4*\cos(\theta_{if})*E_f*E_i+\sin(
      \phi_0 )*(a_2.p_f)*(k.p_f)^2*(k.p_i)^2*c^8-\sin(\phi_0)*(a_2.p_f)*(k.p_f)^2*(k.p_i)*
      (k.s_i)*\lambda_f*\lambda_i*|p_f|*c^6*E_i-\sin(\phi_0)*(a_2.p_f)*(k.p_f)^2*(k.p_i)*(k.s_i)*
      \lambda_f*\lambda_i*|p_i|*c^6*\cos(\theta_{if})*E_f-\sin(\phi_0)*(a_2.p_f)*(k.p_f)*(k.p_i)^3*\lambda_f*\lambda_i
      *|p_f|*|p_i|*c^6-\sin(\phi_0)*(a_2.p_f)*(k.p_f)*(k.p_i)^3*\lambda_f*\lambda_i*c^4*\cos(\theta_{if})*
      E_f*E_i-\sin(\phi_0)*(a_2.p_f)*(k.p_f)*(k.p_i)^3*c^8+\sin(\phi_0)*(a_2.p_f)*(k.p_f)*
      (k.p_i)^2*(k.s_i)*\lambda_f*\lambda_i*|p_f|*c^6*E_i+\sin(\phi_0)*(a_2.p_f)*(k.p_f)*(k.p_i)^2
      *(k.s_i)*\lambda_f*\lambda_i*|p_i|*c^6*\cos(\theta_{if})*E_f-2.0*\sin(\phi_0)*(a_2.p_f)*(k.p_f)*(k.p_i)
      ^2*\lambda_f*\lambda_i*|p_i|^2*c^4*\cos(\theta_{if})*E_f*\omega+2.0*\sin(\phi_0)*(a_2.p_f)*(k.p_f)
      *(k.p_i)^2*\lambda_f*\lambda_i*c^2*\cos(\theta_{if})*E_f*E_i^2*\omega+2.0*\sin(\phi_0)*
      (a_2.p_f)*(k.p_f)*(k.p_i)^2*c^6*E_i*\omega-\sin(\phi_0)*(a_2.p_f)*(k.p_f)*(k.p_i)*(k.s_f)
      *\lambda_f*\lambda_i*(a^2)*|p_i|*c^4*\omega+\sin(\phi_0)*(a_2.p_f)*(k.p_f)*(k.p_i)*\lambda_f*\lambda_i*
      (a^2)*|p_f|*|p_i|*c^2*\omega^2-\sin(\phi_0)*(a_2.p_f)*(k.p_f)*(k.p_i)*\lambda_f*\lambda_i*(a^2)*
      \cos(\theta_{if})*E_f*E_i*\omega^2-\sin(\phi_0)*(a_2.p_f)*(k.p_f)*(k.p_i)*(a^2)*c^4*\omega
      ^2+\sin(\phi_0)*(a_2.p_f)*(k.p_f)*(k.s_f)*(k.s_i)*\lambda_f*\lambda_i*(a^2)*c^4*E_i*\omega-
      \sin(\phi_0)*(a_2.p_f)*(k.p_f)*(k.s_i)*\lambda_f*\lambda_i*(a^2)*|p_f|*c^2*E_i*\omega^2+\sin
      (\phi_0 )*(a_2.p_f)*(k.p_f)*(k.s_i)*\lambda_f*\lambda_i*(a^2)*|p_i|*c^2*\cos(\theta_{if})*E_f*\omega^2+\sin(\phi_0)*(a_2.p_f)*(k.p_i)^2*(k.s_f)*\lambda_f*\lambda_i*(a^2)*|p_i|*c^4*\omega-\sin(
      \phi_0 )*(a_2.p_f)*(k.p_i)*(k.s_f)*(k.s_i)*\lambda_f*\lambda_i*(a^2)*c^4*E_i*\omega-\sin(\phi_0)
      *(a_2.p_i)*(k.p_f)^3*(k.p_i)*\lambda_f*\lambda_i*|p_f|*|p_i|*c^6-\sin(\phi_0)*(a_2.p_i)*(k.p_f)^3
      *(k.p_i)*\lambda_f*\lambda_i*c^4*\cos(\theta_{if})*E_f*E_i-\sin(\phi_0)*(a_2.p_i)*(k.p_f)^3*(k.p_i)*
      c^8+\sin(\phi_0)*(a_2.p_i)*(k.p_f)^2*(k.p_i)^2*\lambda_f*\lambda_i*|p_f|*|p_i|*c^6+\sin(
      \phi_0 )*(a_2.p_i)*(k.p_f)^2*(k.p_i)^2*\lambda_f*\lambda_i*c^4*\cos(\theta_{if})*E_f*E_i+\sin(
      \phi_0 )*(a_2.p_i)*(k.p_f)^2*(k.p_i)^2*c^8+\sin(\phi_0)*(a_2.p_i)*(k.p_f)^2*(k.p_i)*
      (k.s_f)*\lambda_f*\lambda_i*|p_f|*c^6*\cos(\theta_{if})*E_i+\sin(\phi_0)*(a_2.p_i)*(k.p_f)^2*(k.p_i)*
      (k.s_f)*\lambda_f*\lambda_i*|p_i|*c^6*E_f-2.0*\sin(\phi_0)*(a_2.p_i)*(k.p_f)^2*(k.p_i)*\lambda_f*
      \lambda_i*|p_f|^2*c^4*\cos(\theta_{if})*E_i*\omega+2.0*\sin(\phi_0)*(a_2.p_i)*(k.p_f)^2*
      (k.p_i)*\lambda_f*\lambda_i*c^2*\cos(\theta_{if})*E_f^2*E_i*\omega+2.0*\sin(\phi_0)*(a_2.p_i)*
      (k.p_f)^2*(k.p_i)*c^6*E_f*\omega+\sin(\phi_0)*(a_2.p_i)*(k.p_f)^2*(k.s_i)*\lambda_f*\lambda_i
      *(a^2)*|p_f|*c^4*\omega-\sin(\phi_0)*(a_2.p_i)*(k.p_f)*(k.p_i)^2*(k.s_f)*\lambda_f*\lambda_i*
      |p_f|*c^6*\cos(\theta_{if})*E_i-\sin(\phi_0)*(a_2.p_i)*(k.p_f)*(k.p_i)^2*(k.s_f)*\lambda_f*\lambda_i*
      |p_i|*c^6*E_f-\sin(\phi_0)*(a_2.p_i)*(k.p_f)*(k.p_i)*(k.s_i)*\lambda_f*\lambda_i*(a^2)*|p_f|*c^
      4*\omega+\sin(\phi_0)*(a_2.p_i)*(k.p_f)*(k.p_i)*\lambda_f*\lambda_i*(a^2)*|p_f|*|p_i|*c^2*\omega
      ^2-\sin(\phi_0)*(a_2.p_i)*(k.p_f)*(k.p_i)*\lambda_f*\lambda_i*(a^2)*\cos(\theta_{if})*E_f*E_i*\omega^
      2-\sin(\phi_0)*(a_2.p_i)*(k.p_f)*(k.p_i)*(a^2)*c^4*\omega^2-\sin(\phi_0)*(a_2.p_i)*(k.p_f)*(k.s_f)*(k.s_i)*\lambda_f*\lambda_i*(a^2)*c^4*E_f*\omega+
      \sin(\phi_0)*(a_2.p_i)*(k.p_i)*(k.s_f)*(k.s_i)*\lambda_f*\lambda_i*(a^2)*c^4*E_f*\omega+\sin(
      \phi_0 )*(a_2.p_i)*(k.p_i)*(k.s_f)*\lambda_f*\lambda_i*(a^2)*|p_f|*c^2*\cos(\theta_{if})*E_i*\omega^2-
      \sin(\phi_0)*(a_2.p_i)*(k.p_i)*(k.s_f)*\lambda_f*\lambda_i*(a^2)*|p_i|*c^2*E_f*\omega^2-\sin(
      \phi_0 )*(a_2.s_f)*(k.p_f)^2*(k.p_i)^2*\lambda_f*\lambda_i*|p_f|*c^6*\cos(\theta_{if})*E_i-\sin(
      \phi_0 )*(a_2.s_f)*(k.p_f)^2*(k.p_i)^2*\lambda_f*\lambda_i*|p_i|*c^6*E_f+\sin(\phi_0)*(a_2.s_f)
      *(k.p_f)^2*(k.p_i)*(k.s_i)*\lambda_f*\lambda_i*|p_f|*|p_i|*c^8*\cos(\theta_{if})+\sin(\phi_0)*(a_2.s_f)*
      (k.p_f)^2*(k.p_i)*(k.s_i)*\lambda_f*\lambda_i*c^{10}+\sin(\phi_0)*(a_2.s_f)*(k.p_f)^2*(k.p_i)*
      (k.s_i)*\lambda_f*\lambda_i*c^6*E_f*E_i+\sin(\phi_0)*(a_2.s_f)*(k.p_f)^2*(k.p_i)*\lambda_f*\lambda_i
      *(a^2)*|p_i|*c^4*\omega-\sin(\phi_0)*(a_2.s_f)*(k.p_f)^2*(k.s_i)*\lambda_f*\lambda_i*(a^2)*c^
      4*E_i*\omega+\sin(\phi_0)*(a_2.s_f)*(k.p_f)*(k.p_i)^3*\lambda_f*\lambda_i*|p_f|*c^6*\cos(\theta_{if})
      *E_i+\sin(\phi_0)*(a_2.s_f)*(k.p_f)*(k.p_i)^3*\lambda_f*\lambda_i*|p_i|*c^6*E_f-\sin(
      \phi_0 )*(a_2.s_f)*(k.p_f)*(k.p_i)^2*(k.s_i)*\lambda_f*\lambda_i*|p_f|*|p_i|*c^8*\cos(\theta_{if})-\sin(
      \phi_0 )*(a_2.s_f)*(k.p_f)*(k.p_i)^2*(k.s_i)*\lambda_f*\lambda_i*c^{10}-\sin(\phi_0)*(a_2.s_f)*(k.p_f)
      *(k.p_i)^2*(k.s_i)*\lambda_f*\lambda_i*c^6*E_f*E_i-\sin(\phi_0)*(a_2.s_f)*(k.p_f)*(k.p_i)^
      2*\lambda_f*\lambda_i*(a^2)*|p_i|*c^4*\omega+2.0*\sin(\phi_0)*(a_2.s_f)*(k.p_f)*(k.p_i)^2*
      \lambda_f*\lambda_i*|p_f|*|p_i|^2*c^6*\cos(\theta_{if})*\omega-2.0*\sin(\phi_0)*(a_2.s_f)*(k.p_f)*
      (k.p_i)^2*\lambda_f*\lambda_i*|p_f|*c^4*\cos(\theta_{if})*E_i^2*\omega+2.0*\sin(\phi_0)*(a_2.s_f)
      *(k.p_f)*(k.p_i)^2*\lambda_f*\lambda_i*|p_i|*c^8*\omega\sin(\phi_0)*(a_2.s_f)*(k.p_f)*(k.p_i)*(k.s_i)*\lambda_f*\lambda_i*(a^2)*c^4*E_i*\omega+
      \sin(\phi_0)*(a_2.s_f)*(k.p_f)*(k.p_i)*\lambda_f*\lambda_i*(a^2)*|p_f|*c^2*\cos(\theta_{if})*E_i*\omega
      ^2-\sin(\phi_0)*(a_2.s_f)*(k.p_f)*(k.p_i)*\lambda_f*\lambda_i*(a^2)*|p_i|*c^2*E_f*\omega^2-
      \sin(\phi_0)*(a_2.s_f)*(k.p_f)*(k.s_i)*\lambda_f*\lambda_i*(a^2)*|p_f|*|p_i|*c^4*\cos(\theta_{if})*\omega^
      2-\sin(\phi_0)*(a_2.s_f)*(k.p_f)*(k.s_i)*\lambda_f*\lambda_i*(a^2)*c^6*\omega^2+\sin(\phi_0)*
      (a_2.s_f)*(k.p_f)*(k.s_i)*\lambda_f*\lambda_i*(a^2)*c^2*E_f*E_i*\omega^2+\sin(\phi_0)*
      (a_2.s_i)*(k.p_f)^3*(k.p_i)*\lambda_f*\lambda_i*|p_f|*c^6*E_i+\sin(\phi_0)*(a_2.s_i)*(k.p_f)^3
      *(k.p_i)*\lambda_f*\lambda_i*|p_i|*c^6*\cos(\theta_{if})*E_f-\sin(\phi_0)*(a_2.s_i)*(k.p_f)^2*(k.p_i)^
      2*\lambda_f*\lambda_i*|p_f|*c^6*E_i-\sin(\phi_0)*(a_2.s_i)*(k.p_f)^2*(k.p_i)^2*\lambda_f*\lambda_i
      *|p_i|*c^6*\cos(\theta_{if})*E_f-\sin(\phi_0)*(a_2.s_i)*(k.p_f)^2*(k.p_i)*(k.s_f)*\lambda_f*\lambda_i*
      |p_f|*|p_i|*c^8*\cos(\theta_{if})-\sin(\phi_0)*(a_2.s_i)*(k.p_f)^2*(k.p_i)*(k.s_f)*\lambda_f*\lambda_i*c
      ^{10}-\sin(\phi_0)*(a_2.s_i)*(k.p_f)^2*(k.p_i)*(k.s_f)*\lambda_f*\lambda_i*c^6*E_f*E_i-
      \sin(\phi_0)*(a_2.s_i)*(k.p_f)^2*(k.p_i)*\lambda_f*\lambda_i*(a^2)*|p_f|*c^4*\omega+2.0*\sin(
      \phi_0 )*(a_2.s_i)*(k.p_f)^2*(k.p_i)*\lambda_f*\lambda_i*|p_f|^2*|p_i|*c^6*\cos(\theta_{if})*\omega+2.0
      *\sin(\phi_0)*(a_2.s_i)*(k.p_f)^2*(k.p_i)*\lambda_f*\lambda_i*|p_f|*c^8*\omega-2.0*\sin(
      \phi_0 )*(a_2.s_i)*(k.p_f)^2*(k.p_i)*\lambda_f*\lambda_i*|p_i|*c^4*\cos(\theta_{if})*E_f^2*\omega+\sin
      (\phi_0 )*(a_2.s_i)*(k.p_f)*(k.p_i)^2*(k.s_f)*\lambda_f*\lambda_i*|p_f|*|p_i|*c^8*\cos(\theta_{if})+\sin(
      \phi_0 )*(a_2.s_i)*(k.p_f)*(k.p_i)^2*(k.s_f)*\lambda_f*\lambda_i*c^{10}\sin(\phi_0)*(a_2.s_i)*(k.p_f)*(k.p_i)^2*(k.s_f)*\lambda_f*\lambda_i*c^6*E_f*E_i+
      \sin(\phi_0)*(a_2.s_i)*(k.p_f)*(k.p_i)^2*\lambda_f*\lambda_i*(a^2)*|p_f|*c^4*\omega+\sin(\phi_0
      )*(a_2.s_i)*(k.p_f)*(k.p_i)*(k.s_f)*\lambda_f*\lambda_i*(a^2)*c^4*E_f*\omega-\sin(\phi_0)*(a_2.s_i)
      *(k.p_f)*(k.p_i)*\lambda_f*\lambda_i*(a^2)*|p_f|*c^2*E_i*\omega^2+\sin(\phi_0)*(a_2.s_i)*
      (k.p_f)*(k.p_i)*\lambda_f*\lambda_i*(a^2)*|p_i|*c^2*\cos(\theta_{if})*E_f*\omega^2-\sin(\phi_0)*
      (a_2.s_i)*(k.p_i)^2*(k.s_f)*\lambda_f*\lambda_i*(a^2)*c^4*E_f*\omega-\sin(\phi_0)*(a_2.s_i)*
      (k.p_i)*(k.s_f)*\lambda_f*\lambda_i*(a^2)*|p_f|*|p_i|*c^4*\cos(\theta_{if})*\omega^2-\sin(\phi_0)*
      (a_2.s_i)*(k.p_i)*(k.s_f)*\lambda_f*\lambda_i*(a^2)*c^6*\omega^2+\sin(\phi_0)*(a_2.s_i)*(k.p_i)*
      (k.s_f)*\lambda_f*\lambda_i*(a^2)*c^2*E_f*E_i*\omega^2-\sin(\phi_0)*(k.p_f)^2*(k.p_i)*
      \lambda_f*|p_f|^2*c^7*\omega*\sin(\varphi_f)*\sin(\theta_f)*|a|+\sin(\phi_0)*(k.p_f)^2*(k.p_i)*
      \lambda_f*|p_f|*|p_i|*c^7*\omega*\sin(\varphi_i)*\sin(\theta_i)*|a|+\sin(\phi_0)*(k.p_f)^2*(k.p_i)
      *\lambda_f*|p_i|*c^6*\cos(\theta_f)*E_f*\omega*\sin(\varphi_i)*\sin(\theta_i)*|a|-\sin(\phi_0)*(k.p_f)
      ^2*(k.p_i)*\lambda_f*|p_i|*c^6*\cos(\theta_i)*E_f*\omega*\sin(\varphi_f)*\sin(\theta_f)*|a|+\sin(
      \phi_0 )*(k.p_f)^2*(k.p_i)*\lambda_f*c^5*E_f^2*\omega*\sin(\varphi_f)*\sin(\theta_f)*|a|+\sin(
      \phi_0 )*(k.p_f)^2*(k.p_i)*\lambda_f*c^5*E_f*E_i*\omega*\sin(\varphi_f)*\sin(\theta_f)*|a|+
      \sin(\phi_0)*(k.p_f)^2*(k.p_i)*\lambda_i*|p_f|*|p_i|*c^7*\omega*\sin(\varphi_f)*\sin(\theta_f)*|a|
      +\sin(\phi_0)*(k.p_f)^2*(k.p_i)*\lambda_i*|p_f|*c^6*\cos(\theta_f)*E_i*\omega*\sin(\varphi_i)*
      \sin(\theta_i)*|a|-\sin(\phi_0)*(k.p_f)^2*(k.p_i)*\lambda_i*|p_f|*c^6*\cos(\theta_i)*E_i*
      \omega*\sin(\varphi_f)*\sin(\theta_f)*|a|-\sin(\phi_0)*(k.p_f)^2*(k.p_i)*\lambda_i*|p_i|^2*c^7*\omega*\sin(\varphi_i)*\sin(\theta_i)*
      |a|+\sin(\phi_0)*(k.p_f)^2*(k.p_i)*\lambda_i*c^5*E_f*E_i*\omega*\sin(\varphi_i)*
      \sin(\theta_i)*|a|+\sin(\phi_0)*(k.p_f)^2*(k.p_i)*\lambda_i*c^5*E_i^2*\omega*\sin(\varphi_i)
      *\sin(\theta_i)*|a|+\sin(\phi_0)*(k.p_f)*(k.p_i)^2*\lambda_f*|p_f|^2*c^7*\omega*\sin(\varphi_f)
      *\sin(\theta_f)*|a|-\sin(\phi_0)*(k.p_f)*(k.p_i)^2*\lambda_f*|p_f|*|p_i|*c^7*\omega*
      \sin(\varphi_i)*\sin(\theta_i)*|a|+\sin(\phi_0)*(k.p_f)*(k.p_i)^2*\lambda_f*|p_i|*c^6*\cos(\theta_f)*
      E_f*\omega*\sin(\varphi_i)*\sin(\theta_i)*|a|-\sin(\phi_0)*(k.p_f)*(k.p_i)^2*\lambda_f*|p_i|*c^6
      *\cos(\theta_i)*E_f*\omega*\sin(\varphi_f)*\sin(\theta_f)*|a|-\sin(\phi_0)*(k.p_f)*(k.p_i)^2*\lambda_f*c
      ^5*E_f^2*\omega*\sin(\varphi_f)*\sin(\theta_f)*|a|-\sin(\phi_0)*(k.p_f)*(k.p_i)^2*\lambda_f*c
      ^5*E_f*E_i*\omega*\sin(\varphi_f)*\sin(\theta_f)*|a|-\sin(\phi_0)*(k.p_f)*(k.p_i)^2*\lambda_i
      *|p_f|*|p_i|*c^7*\omega*\sin(\varphi_f)*\sin(\theta_f)*|a|+\sin(\phi_0)*(k.p_f)*(k.p_i)^2*
      \lambda_i*|p_f|*c^6*\cos(\theta_f)*E_i*\omega*\sin(\varphi_i)*\sin(\theta_i)*|a|-\sin(\phi_0)*(k.p_f)*
      (k.p_i)^2*\lambda_i*|p_f|*c^6*\cos(\theta_i)*E_i*\omega*\sin(\varphi_f)*\sin(\theta_f)*|a|+\sin(
      \phi_0 )*(k.p_f)*(k.p_i)^2*\lambda_i*|p_i|^2*c^7*\omega*\sin(\varphi_i)*\sin(\theta_i)*|a|-\sin(
      \phi_0 )*(k.p_f)*(k.p_i)^2*\lambda_i*c^5*E_f*E_i*\omega*\sin(\varphi_i)*\sin(\theta_i)*|a|-
      \sin(\phi_0)*(k.p_f)*(k.p_i)^2*\lambda_i*c^5*E_i^2*\omega*\sin(\varphi_i)*\sin(\theta_i)*|a|
      -\sin(\phi_0)*(k.p_f)*(k.p_i)*\lambda_f*(a^2)*c^3*E_f*\omega^2*\sin(\varphi_f)*\sin(\theta_f)*
      |a|-\sin(\phi_0)*(k.p_f)*(k.p_i)*\lambda_i*(a^2)*c^3*E_i*\omega^2*\sin(\varphi_i)*
      \sin(\theta_i)*|a|-\sin(\phi_0)*(k.p_f)*(k.s_i)*\lambda_i*(a^2)*|p_f|*c^5*\omega^2*\sin(\varphi_f)*\sin(\theta_f)
      *|a|+\sin(\phi_0)*(k.p_f)*(k.s_i)*\lambda_i*(a^2)*|p_i|*c^5*\omega^2*\sin(\varphi_i)*
      \sin(\theta_i)*|a|+\sin(\phi_0)*(k.p_i)^2*\lambda_f*(a^2)*c^3*E_f*\omega^2*\sin(\varphi_f)*
      \sin(\theta_f)*|a|+\sin(\phi_0)*(k.p_i)^2*\lambda_i*(a^2)*c^3*E_i*\omega^2*\sin(\varphi_i)*
      \sin(\theta_i)*|a|+\sin(\phi_0)*(k.p_i)*(k.s_i)*\lambda_i*(a^2)*|p_f|*c^5*\omega^2*\sin(\varphi_f)
      *\sin(\theta_f)*|a|-\sin(\phi_0)*(k.p_i)*(k.s_i)*\lambda_i*(a^2)*|p_i|*c^5*\omega^2*
      \sin(\varphi_i)*\sin(\theta_i)*|a|-\sin(\phi_0)*(k.p_i)*\lambda_f*(a^2)*|p_f|^2*c^3*\omega^3*
      \sin(\varphi_f)*\sin(\theta_f)*|a|+\sin(\phi_0)*(k.p_i)*\lambda_f*(a^2)*|p_f|*|p_i|*c^3*\omega^3*
      \sin(\varphi_i)*\sin(\theta_i)*|a|-\sin(\phi_0)*(k.p_i)*\lambda_f*(a^2)*|p_i|*c^2*\cos(\theta_f)*E_f*
      \omega^3*\sin(\varphi_i)*\sin(\theta_i)*|a|+\sin(\phi_0)*(k.p_i)*\lambda_f*(a^2)*|p_i|*c^2*
      \cos(\theta_i)*E_f*\omega^3*\sin(\varphi_f)*\sin(\theta_f)*|a|+\sin(\phi_0)*(k.p_i)*\lambda_f*(a^2)*c*
      E_f^2*\omega^3*\sin(\varphi_f)*\sin(\theta_f)*|a|-\sin(\phi_0)*(k.p_i)*\lambda_f*(a^2)*c*E_f*
      E_i*\omega^3*\sin(\varphi_f)*\sin(\theta_f)*|a|-\sin(\phi_0)*(k.p_i)*\lambda_i*(a^2)*|p_f|*|p_i|*
      c^3*\omega^3*\sin(\varphi_f)*\sin(\theta_f)*|a|-\sin(\phi_0)*(k.p_i)*\lambda_i*(a^2)*|p_f|*c^2
      *\cos(\theta_f)*E_i*\omega^3*\sin(\varphi_i)*\sin(\theta_i)*|a|+\sin(\phi_0)*(k.p_i)*\lambda_i*(a^2)*
      |p_f|*c^2*\cos(\theta_i)*E_i*\omega^3*\sin(\varphi_f)*\sin(\theta_f)*|a|+\sin(\phi_0)*(k.p_i)*
      \lambda_i*(a^2)*|p_i|^2*c^3*\omega^3*\sin(\varphi_i)*\sin(\theta_i)*|a|+\sin(\phi_0)*(k.p_i)*
      \lambda_i*(a^2)*c*E_f*E_i*\omega^3*\sin(\varphi_i)*\sin(\theta_i)*|a|-\sin(\phi_0)*(k.p_i)*
      \lambda_i*(a^2)*c*E_i^2*\omega^3*\sin(\varphi_i)*\sin(\theta_i)*|a|\Big\}/[2.0*(k.p_f)^2*(k.p_i)^2*c^7] $
\end{sloppypar}
\end{widetext}

\end{document}